\documentclass[journal,onecolumn]{IEEEtran}
\usepackage{xcolor}
\usepackage{amsmath}
\usepackage{amssymb}
\usepackage{physics}
\usepackage{array}
\usepackage{tabu}
\usepackage{layout}
\usepackage{arydshln}
\usepackage{booktabs,caption}
\usepackage[flushleft]{threeparttable}
\newtheorem{theorem}{Theorem}
\newtheorem{lemma}{Lemma}
\newtheorem{definition}{Definition}

\newtheorem{corollary}{Corollary}
\newtheorem{remark}{Remark}

\usepackage{csquotes}
\usepackage{bbm}
\usepackage{mathtools}
\usepackage{comment}
\usepackage[makeroom]{cancel}
\DeclarePairedDelimiter{\ceil}{\lceil}{\rceil}
\definecolor{cool_green}{rgb}{0.0, 0.5, 0.0}

\usepackage[mathscr]{euscript}
\usepackage{dsfont}

\renewcommand{\Pr}{{\operatorname{Pr}}}
\renewcommand{\Tr}{{\operatorname{Tr}}}

\newcommand{\ma}{{\operatorname{max}}}
\newcommand{\h}{{\operatorname{H}}}
\newcommand{\Pu}{{\operatorname{P}}}

\newcommand{\cB}{{\mathcal{B}^{\varepsilon}}}
\newcommand{\crho}{{\rho^{XAB}=\sum_{x}p(x)\ketbra{x}\otimes\rho^{AB}_{x}}}
\newcommand{\ct}{{\big\|}}

\begin{document}
\title{Publicness, Privacy and Confidentiality in the Single-Serving Quantum Broadcast Channel}
\author{Farzin~Salek,~\IEEEmembership{Student Member,~IEEE,} Min-Hsiu~Hsieh,~\IEEEmembership{Senior Member,~IEEE,} Javier~R.~Fonollosa,~\IEEEmembership{Senior Member,~IEEE,}
\thanks{F. Salek and J. R. Fonollosa are with the Department
of Signal Theory and Communications, Universitat Polit\`{e}cnica de Catalunya, Barcelona,
Spain, e-mail: (farzin.salek@upc.edu, javier.fonollosa@upc.edu).}
\thanks{Min-Hsiu Hsieh is with the Center for Quantum Software and Information, Sydney University of Technology, Sydney, Australia, e-mail: (min-hsiu.hsieh@uts.edu.au).}}

\maketitle
\section*{Abstract}
The 2-receiver broadcast channel is studied: a network with three parties where the transmitter and one of the receivers are the primarily involved parties and the other receiver considered as third party. The messages that are determined to be communicated are classified into public, private and confidential based on the information they convey. The public message contains information intended for both parties and is required to be decoded correctly by both of them, the private message is intended for the primary party only, however, there is no secrecy requirement imposed upon it meaning that it can possibly be exposed to the third party and finally the confidential message containing information intended exclusively for the primary party such that this information must be kept completely secret from the other receiver. A trade-off arises between the rates of the three messages, when one of the rates is high, the other rates may need to be reduced to guarantee the reliable transmission of all three messages. The encoder performs the necessary equivocation by virtue of dummy random numbers whose rate is assumed to be limited and should be considered in the trade-off as well. 
We study this trade-off in the one-shot regime of a quantum broadcast channel by providing achievability and (weak) converse regions. In the achievability, we prove and use a conditional version of the convex-split lemma as well as position-based decoding. By studying the asymptotic behaviour of our bounds, we will recover several well-known asymptotic results in the literature.
%\footnote{In the literature, the third-party receiver is also called adversary, eavesdropper, malicious receiver, etc. Although the third party is subject to reliably receiving some information, those names still seem reasonable since it is yet an adversary with respect to some other information.}

% In the context of the classical information theory, the problem of communication over a discrete memoryless channel with a secrecy requirement was studied under the name of the (degraded) wiretap channel: a sender tries to communicate as much reliable and confidential messages as possible to a legitimate receiver who enjoys some advantage over an eavesdropper. This model was latter generalized by dropping the advantage of the legitimate receiver and including common messages. Latter researchers considered this problem with a constraint on the rate of the dummy randomness needed for obfuscation. In the context of quantum information theory, the problem of confidential transmission of messages is studied in both asymptotic i.i.d and one-shot regimes.

\section{Introduction}
Consider a communication model in which a sender attempts to reliably transmit a message to a receiver while hiding it from an eavesdropper. This model was introduced and studied by Wyner under the name of the wiretap channel \cite{Wyner}. The basic idea underlying the coding scheme of Wyner is to generate a certain number of sequences and partition them into bins which are labeled with the messages to be transmitted. To send a message, a sequence from the message bin is randomly selected and transmitted. In the original model of Wyner, the eavesdropper is put at a physical disadvantage with respect to the (legitimate) receiver meaning that upon transmission over the channel, the eavesdropper only receives a noisy version of the information received by the receiver (for this reason his model is usually referred to as the degraded wiretap channel).
The degraded wiretap channel model was  latter enhanced by Csisz\'ar and K\"orner \cite{Csiszar-Korner} by introducing a public (or common) message that is piggybacked on top of the confidential message and is supposed to be reliably decoded by both the receiver and the eavesdropper. Furthermore, in this new model called broadcast channel with confidential messages (BCC), the receiver has no advantage over the eavesdropper. The coding scheme of BCC consists of superposition coding \cite {Coverbook} to encode the confidential message on top of the common message and Wyner's codebook structure with local randomness for equivocation. The coding scheme of BCC consists of superposition coding \cite {Coverbook} to encode the confidential message on top of the common message and Wyner's codebook structure with local randomness for equivocation. The most important contribution of BCC is prepending a prefixing stochastic map to the channel and using a then-new single-letterization trick in the converse proof. 

%Let $V,X,Y$ and $Z$ be random variables distributed as $p(v,x,y,z)=p(v)p(x|v)p(y,z|x)$, where $p(y,z|x)$ models the classical channel such that for an input $x$, the receiver and the eavesdropper receive $y$ and $z$, respectively. Then, the rate of the confidential message (known as the secrecy capacity) is as follows \cite{Csiszar-Korner}: 
%\begin{align}
%\label{fullsecrecy}
%C_{s}\coloneqq \max_{p(v,x)}\left[ I(V;Y)-I(V;Z)\right].
%\end{align}

The simulation of the prefixing stochastic map is performed from random numbers using some method such as the channel simulation \cite{Steinberg-Verdu}. Therefore, we can conclude that at two points the BCC uses random numbers, first in selecting codewords randomly from the bins and second simulating the prefixing map. Nonetheless, in the original works of Wyner and Csisz\'r-K\"orner, the encoder was assumed to have an unlimited amount of randomness at its disposal and a detailed analysis reveals that once there is a constraint on the amount of the randomness, the original works cannot guarantee the secrecy.

Latter in \cite{csiszar-korner-book}, Csisz\'ar and K\"orner proposed an alternative description for the BCC  such that the message to be transmitted consists of two \textit{independent} parts, a confidential part defined in the same sense as the original BCC and a non-secret part, i.e, a message without any secrecy requirement placed on it. The striking difference between two descriptions is that in the original version, no message by no means was allowed to be overheard by the eavesdropper without jeopardizing the secrecy, while in the alternative version, some non-secret message is allowed to be potentially intercepted by the eavesdropper without compromising the secrecy. To put another way, the alternative version allows for substituting (maybe part of) the local randomness by some non-secret, or private message. Although the alternative does not provide clues as to the secrecy under the absence of the unlimited randomness, it triggers the idea that some private message can play the role of the randomness if necessary.

In \cite{Bloch-Kliewer}, Bloch and Kliewer studied the degraded wiretap channel when the randomness is constrained and not necessarily uniform. The general BCC model with rate-constrained randomness was studied by Watanabe and Oohama  in \cite{Wat-Ooh15}. In this paper, the trade-off between the private message and the dummy randomness was recognized for classical channels. To obtain the so-called trade-off, they have used a superposition scheme to replace the prefixing stochastic map proposed originally by Chia and El Gamal in studying the 3-receiver broadcast channel \cite{Chia-ElGamal12}. The idea of Chia and El Gamal was to replace the prefixing stochastic map with a deterministic codebook whose codewords are selected randomly. It is investigated in \cite{Wat-Ooh15} that the amount of the randomness needed to select a codeword randomly following Chia-El Gamal scheme is less that the randomness needed to simulate the prefixing map. This shows that the direct concatenation of ordinary random encoding and channel prefixing with channel simulation is in general suboptimal.

The quantum wiretap channel was studied by Cai-Winter-Yeung \cite{Cai-Winter-Yeung} and Devetak \cite{Dev-private} and the capacity is given by a regularized formula meaning that unlike its classical counterpart, the capacity of the quantum wiretap channel is not completely understood. The ability of quantum channels to preserve quantum superpositions gives rise to purely quantum information-processing tasks that there are no classical counterparts for them. The quantum capacity, i.e., the ability of a quantum channel to transmit qubits, is one such example. The ability of quantum channels to convey both classical and quantum information made Devetak and Shor to unify two tasks and study the simultaneously achievable transmission rates of classical and quantum information \cite{Devetak-Shor}. Their protocol is conceptually related to the superposition coding where for each classical message a different quantum code is used and the capacity region is given in form of a regularization of some single-letter region.

Lacking unlimited resources such as many instances of channels or many copies of certain states in nature, triggered a new area of research called information theory with finite resources. This area has been drawn significant attention over the past years, see \cite{Marcobook} for a survey. The extreme scenario where only one instance of a certain resource such as a channel use or a source state is available, is generally called \textit{one-shot} regime and such a channel (res. source) is called \textit{single-serving} channel (res. source). One-shot channel model is the most general model and its capacity to accomplish several information-processing tasks have been studied. The question of the number of bits that can be transmitted with an error of at most $\varepsilon>0$ by a single use of a classical channel is answered in \cite{Renner-single-serving} where the capacity is characterized in terms of smooth min- and max-entropies. The same question for the quantum channels is studied in \cite{Milan-Nilanjana} following a hypothesis-testing approach and the capacity is characterized in terms of general R\'eyni entropies. 

A novel positive-operator valued measurement (POVM) is introduced in \cite{Wang-Renner} yielding an achievability bound for the capacity of the classical-quantum (cq) channels. The POVM construction as well as the converse proof follow a hypothesis-testing procedure and the result is governed by a smooth relative entropy quantity. This result was rederived in \cite{Mark2017} by deploying a coding scheme known as the position-based decoding \cite{ANR2017}. While the position-based decoding ensures the reliability of the transmitted messages,  \cite{Mark2017} employed another tool called convex-split lemma \cite{Anurag-convex-split} guaranteeing that a malicious third party having partial information about the messages cannot be able to crack them if certain condition holds resulting in a capacity theorem for the one-shot wiretap quantum channel. Position-based decoding and convex-split lemma are governed by quantities known as smooth relative entropies which will be defined in the next section. One can think of the position-based decoding and convex-split lemma as a packing and a covering lemma, respectively. Another result on the one-shot capacity of the quantum wiretap channel was given by  \cite{JPN2017}. In this work, the reliability of the messages are ensured by employing the POVMs introduced in \cite{Wang-Renner} and the confidentiality of the messages is established by proving a novel one-shot covering lemma analogous in approach to \cite{Covering.A.W.2002}.

From a different perspective, \cite{Renes-Renner11} showed that two primitive information-theoretic protocols namely information reconciliation and privacy amplification are capable of directly constructing optimal two-terminal protocols. The appealing feature of this approach is that the primitive protocols are used to build up perhaps more complicated schemes in such a way that the internal workings of the primitives themselves are not of concern (much like concatenation of a source code and a channel code to perform joint source-channel coding). This approach yields achievability bounds for the public and confidential capacities of cq channels and their tightness also established by proving corresponding converse bounds. The quantum capacity of a quantum channel for one or a finite number of uses is studied in \cite{Buscemi-Datta-2010}. The current authors with their colleagues in a former work \cite{Farzin2018}, unified the problems of one-shot transmission of public and confidential information over quantum channels and proposed a protocol for simultaneously achieviable public and confidential rates as well as tight converse bounds. Latter, following the Devetak's proof of the quantum capacity \cite{Dev-private}, they proved a one-shot result for simultaneous transmission of the classical and quantum information \cite{Farzin.arXiv.2018}. 

In this work we aim to study the problem of transmission of common, private and confidential messages with randomness constrained encoder over a single use of a 2-receiver quantum broadcast channel. This problem in the asymptotic setting of a memoryless classical channel was studied in \cite{Wat-Ooh15}. One technical contribution of \cite{Wat-Ooh15} is the study of the channel resolvability problem via superposition of classical codewords. The quantum channel resolvability via superpositions in the one-shot regime is studied in \cite{Anurag-Hayashi-Warsi.2018} in the context of the Galfand-Pinsker quantum wiretap channel. Our technical tools in achievability are position-based decoding and convex-split lemma. The setup of our problem requires a new notion of the position-based decoding and convex-split lemma, where we call them conditional position-based decoding and conditional convex-split lemma. The former leads to an operational interpretation of a recently-defined mutual information-like quantity and the latter, which is proved and should be considered as an independent lemma on its own right, gives rise to a new mutual-information like quantity as well as its operational meaning. We note that in a former work of the current authors \cite{Farzin.arXiv.2018}, different definitions and approaches was taken to address the problem. We believe that our definitions in this paper are more reasonable. The broad scope of the rate region developed in this paper enables us to recover not only the classical result of Watanabe and Oohama \cite{Wat-Ooh15}, but also the case of simultaneous transmission of public and private information \cite{Farzin2018}, the simultaneous transmission of the classical and quantum information \cite{Devetak-Shor}, \cite{Farzin.arXiv.2018} and the capacity region of the quantum broadcast channel by Yard-Hayden-Devetak \cite{6034754}.

The rest of the paper is organized as follows. We start with miscellaneous definitions in section II. Section III is devoted to the description of the information-processing task, the definition of the code for the task and our main results. We prove our achievability region in section IV and our converse region in section V. The asymptotic analysis is provided in section VI. We finally conclude the paper in section VII. The proof of the conditional convex-split lemma as well as several other lemmas are given in the appendix.

\section{Miscellaneous Definitions}
We use the following conventions throughout the paper. The capital letters $X,Y$, etc. will denote random variables whose realizations and the alphabets will be shown by the corresponding small and calligraphic letters, respectively. The classical systems associated to the random variables will be denoted by the same capital letters. Quantum systems $A, B$, etc. are associated with (finite dimensional) Hilbert spaces $\mathcal{H}^{A}, \mathcal{H}^{B}$, etc. The set of positive semi-definite operators acting on $\mathcal{H}$ is denoted by $\mathcal{P}(\mathcal{H})$. Multipartite systems are described by tensor product spaces which we denote by the short notation $\mathcal{H}^{AB...D}=\mathcal{H}^{A}\otimes\mathcal{H}^{B}\otimes...\otimes\mathcal{H}^{D}$. We identify states with their density operators and use superscripts to denote the systems on which the mathematical objects are defined. For example if $\rho^{AB}\in\mathcal{H}^{AB}$, then $\rho^{A}=\Tr_{B}\rho^{AB}$ is implicitly defined as its marginal on $A$. The identity operator on $\mathcal{H}^{A}$ is denoted by $\mathbbm{1}^{A}$.

Denoted by $\mathcal{N}^{A\rightarrow B}$, a quantum channel is a completely positive-trace preserving (CPTP) liner map taking input states from the Hilbert space $\mathcal{H}_{A}$ to output states living in the Hilbert space $\mathcal{H}_{B}$. A quantum broadcast channel $\mathcal{N}^{A\rightarrow BC}$ though, refers to a quantum channel with a single input and two outputs such that when the transmitter inputs a quantum state living in $\mathcal{H}_{A}$, one receiver obtains a state in system $B$ living in $\mathcal{H}^{B}$ while the other receiver obtains system $C$ living in $\mathcal{H}^{C}$. Throughout we assume the receiver obtaining $B$ system is the primary receiver and the other receiver obtaining $C$ is third party. It is also useful to personify the users of the channel such that Alice is the user controlling the input and Bob and Charlie are the recipients of the systems $B$ and $C$, respectively. According to the Stinespring dilation of the CPTP map $\mathcal{N}^{A\rightarrow BC}$ (see for example \cite{Markbook}), there exists an \textit{inaccessible environment} $F$ living in $\mathcal{H}^{F}$ and a unitary operator $U$ acting on $A, B$ and $F$ systems such that
\begin{align}
\label{stinespring}
\mathcal{N}^{A\rightarrow BC}=\Tr_{F}\{U(\rho^{A}\otimes\sigma^{C}\otimes\omega^{F})U^{\dagger}\},
\end{align}
where $\rho^{A}$ is the input state and $\sigma^{C}$ and $\omega^{F}$ are some constant states on systems $C$ and $F$, respectively\footnote{This can be equivalently shown via isometric extension of the channel as $\mathcal{N}^{A\rightarrow BC}(\rho^{A})=\Tr_{F}\{V\rho^{A}V^{\dagger}\}$ where $\mbox{\normalsize V}^{A\rightarrow BCF}_{\mathcal{N}}$ is an isometric extension of the channel.}. An additional trace over $C$ system gives the quantum channel from Alice to Bob $\mathcal{N}^{A\rightarrow B}$ implying that the composite system $E\coloneqq CF$ plays the role of an inaccessible environment for $\mathcal{N}^{A\rightarrow B}$. This should not concern us since we transmit classical information and every classical letter transmitted over the channel can be intercepted by more than one receiver (no violation of no-cloning). However, when it comes to the transmission of the quantum information from Alice to Bob, the $E$ system as a whole is considered the environment surrounding $\mathcal{N}^{A\rightarrow B}$.

The von Neumann entropy and the quantum relative entropy are defined as $S(\rho)\coloneqq-\Tr \rho \log \rho$ and $D(\rho\|\sigma)\coloneqq\Tr(\rho\log\rho-\rho\log\sigma)$, respectively (throughout this paper,
log denotes by default the binary logarithm, and its inverse function exp, unless otherwise stated, is also to
basis 2). Conditional entropy and conditional mutual information, $S(A|B)_{\rho}$ and $I(A;B|C)_{\rho}$, respectively,
are defined in the same way as their classical counterparts:
\begin{align*}
S(A|B)_{\rho} &\coloneqq S(AB)_{\rho} - S(B)_{\rho}, \quad\text{and}\\
I(A ; B|C)ρ &\coloneqq S(A|C)_{\rho} - S(A|BC)_{\rho} = S(AC)_{\rho} + S(BC)_{\rho} - S(ABC)_{\rho} - S(C)_{\rho}.
\end{align*}
The von Neumann entropy and the mutual information can be defined as special cases of the quantum relative entropy; for instance it can be seen that $D(\rho^{AB}\|\rho^{A}\otimes\rho^{B})=I(A;B)_{\rho}$.

The trace distance between two states $\rho$ and $\sigma$ is given as $\frac{1}{2}\|\rho-\sigma\|_{1}$ and the fidelity between them is defined as:
\begin{align*}
F(\rho,\sigma)\coloneqq\|\sqrt{\rho}\sqrt{\sigma}\|_{1}=\Tr\sqrt{\rho^{\frac{1}{2}}\sigma\rho^{\frac{1}{2}}}.
\end{align*}
The fidelity relates to the quantum relative entropy in the following way (Pinsker's inequality) \cite{Csiszar67}:
\begin{align}
\label{Pin}
F^{2}(\rho,\sigma)\geq 2^{-D(\rho\|\sigma)}.
\end{align}
The trace distance (res. fidelity) is a convex (res. concave) functions. Notice the following, for cq states $\rho^{XA}=\sum_{x}p(x)\ketbra{x}\otimes\rho^{A}_{x}$ and $\sigma^{XA}=\sum_{x}p(x)\ketbra{x}\otimes\sigma^{A}_{x}$, we have:
\begin{align}
\label{derandomizationaa}
\frac{1}{2}\big\|\rho^{A}-\sigma^{A}\big\|_{1}\leq\frac{1}{2}\big\|\rho^{XA}-\sigma^{XA}\big\|_{1}=\sum_{x}p(x)\frac{1}{2}\big\|\rho^{A}_{x}-\sigma^{A}_{x}\big\|_{1}.
\end{align}
The fidelity is used to define the purified distance as follows:
\begin{align*}
\Pu(\rho,\sigma)\coloneqq \sqrt{1-F^{2}(\rho,\sigma)}.
\end{align*}
It relates to the trace distance in the following way:
\begin{align*}
\frac{1}{2}\|\rho-\sigma\|_{1}\leq\Pu(\rho,\sigma)\leq\sqrt{\|\rho-\sigma\|_{1}}.
\end{align*}
The purified distance is used to define an $\varepsilon$-ball around a state $\rho$: $\rho'\in\cB(\rho)$ if $\Pu(\rho',\rho)\leq\varepsilon$. The purified distance enjoys several properties similar to those of the trace distance, we list some of them below.
\begin{lemma}[see for example \cite{20.500.11850/153605}]
\begin{itemize}
\item Monotonicity: For quantum states $\rho, \sigma$ and any CPTP map $\mathcal{E}$,
\begin{align*}
\Pu(\mathcal{E}(\rho),\mathcal{E}(\sigma))\leq\Pu(\rho,\sigma).
\end{align*}
\item Triangle inequality: For quantum states $\rho,\sigma$ and $\omega$, it holds that
\begin{align*}
\Pu(\rho,\sigma)\leq\Pu(\rho,\omega)+\Pu(\omega,\sigma).
\end{align*} 
\item Invariance with respect to tensor product states: For quantum states $\rho,\sigma$ and $\omega$, it holds that:
\begin{align*}
\Pu(\rho\otimes\omega,\sigma\otimes\omega)=\Pu(\rho,\sigma).
\end{align*}
The following can also be easily verified:
\begin{align*}
\Pu\big(\sum_{x}p(x)\ketbra{x}\otimes\rho^{A}_{x}\otimes\omega^{B}_{x},\sum_{x}q(x)\ketbra{x}\otimes\sigma^{A}_{x}\otimes\omega^{B}_{x}\big)=\Pu\big(\sum_{x}p(x)\ketbra{x}\otimes\rho^{A}_{x},\sum_{x}q(x)\ketbra{x}\otimes\sigma^{A}_{x}\big).
\end{align*}
\end{itemize}
\end{lemma}

\begin{lemma}[Lemma 17 in \cite{5961850}]
\label{pur}
Let $\rho\in\mathcal{H}$ and $\Pi$ a projector on $\mathcal{H}$, then
\begin{align*}
\Pu(\rho,\Pi\rho\Pi)\leq\sqrt{2\Tr\rho\Pi_{\perp}-(\Tr\rho\Pi_{\perp})^{2}},
\end{align*}
where $\Pi_{\perp}=\mathbbm{1}-\Pi$.
\end{lemma}
\begin{lemma}[corollary 16 in \cite{5961850}]
Let $\rho^{AB}=\ketbra{\varphi}^{AB}\in\mathcal{P}(\mathcal{H}^{AB})$ be a pure state, $\rho^{A}=\Tr_{B}\rho^{AB}$, $\rho^{B}=\Tr_{A}\rho^{AB}$ and let $\Pi^{A}\in\mathcal{P}(\mathcal{H}^{A})$ be a projector in supp$(\rho^{A})$. Then, there exists a dual projector $\Pi^{B}$ on $\mathcal{H}^{B}$ such that
\begin{align*}
(\Pi^{A}\otimes(\rho^{B})^{-\frac{1}{2}})\ket{\varphi}^{AB}=((\rho^{A})^{-\frac{1}{2}}\otimes\Pi^{B})\ket{\varphi}^{AB}.
\end{align*}
\end{lemma}
\begin{lemma}[\cite{20.500.11850/153605}]
\label{equt}
Let $\rho, \sigma\in\mathcal{P}(\mathcal{H})$, then
\begin{itemize}
\item For any $\omega\geq\rho$,
\begin{align*}
\big\|\sqrt{\omega}\sqrt{\sigma}\big\|_{1}\geq\big\|\sqrt{\rho}\sqrt{\sigma}\big\|_{1}.
\end{align*}
\item For any projector $\Pi\in\mathcal{P}(\mathcal{H})$,
\begin{align*}
\big\|\sqrt{\Pi\rho\Pi}\sqrt{\sigma}\big\|_{1}&=\big\|\sqrt{\rho}\sqrt{\Pi\sigma\Pi}\big\|_{1}\\
&=\big\|\sqrt{\Pi\rho\Pi}\sqrt{\Pi\sigma\Pi}\big\|_{1}.
\end{align*}
\end{itemize}
\end{lemma}

\begin{definition}[Hypothesis testing relative entropy \cite{Wang-Renner},\cite{Buscemi-Datta-2010}]
\label{htmi}
Let  $\{\Lambda,\mathbbm{1}-\Lambda\}$ be the elements of a POVM that distinguishes between quantum states $\rho$ and $\sigma$ such that the probability of a correct guess on input $\rho$ equals $\Tr\Lambda\rho$ and a wrong guess on $\sigma$ is made with probability $\Tr\Lambda\sigma$. Let $\varepsilon\in(0,1)$. Then, the hypothesis testing relative entropy is defined as follows:
\begin{align*}
D_{\h}^{\varepsilon}(\rho\|\sigma)\coloneqq\max{\{-\log_{2}\Tr\Lambda\sigma:0\leq\Lambda\leq \mathbbm{1}\wedge\Tr\Lambda\rho\geq1-\varepsilon\}}.
\end{align*}
From the definition above, the hypothesis testing mutual information for a bipartite state $\rho^{AB}$ is defined as follows:
\begin{align*}
I_{\h}^{\varepsilon}(A;B)_{\rho}\coloneqq D_{\h}^{\varepsilon}(\rho^{AB}\|\rho^{A}\otimes\rho^{B}).
\end{align*}
\end{definition}
\begin{lemma}[\cite{Wang-Renner}]
\label{hrelative}
For quantum states $\rho$ and $\sigma$ and a parameter $\varepsilon\in(0,1)$, the following relation exists between the hypothesis testing relative entropy and the quantum relative entropy:
\begin{align*}
D_{\h}^{\varepsilon}(\rho\|\sigma)\leq\frac{1}{1-\varepsilon}(D(\rho\|\sigma)+h_{b}(\varepsilon)),
\end{align*}
where $h_{b}(\varepsilon)\coloneqq-\varepsilon\log\varepsilon-(1-\varepsilon)\log(1-\varepsilon)$ is the binary entropy function. The following is a simple consequence of this lemma. For a bipartite state $\rho^{AB}\in\mathcal{H}^{AB}$, we have
\begin{align}
\label{relationH}
I_{\h}^{\varepsilon}(A;B)_{\rho}\leq\frac{1}{1-\varepsilon}(I(A;B)_{\rho}+h_{b}(\varepsilon)).
\end{align}
\end{lemma}
\begin{definition}[Hypothesis testing conditional mutual information \cite{2018arXiv180607276S}]
\label{htcma}
Let $\rho^{XAB}\coloneqq\sum_{x}p(x)|x\rangle\langle x|^{X}\otimes\rho^{AB}_{x}$, $\rho^{A-X-B}\coloneqq\sum_{x}p(x)|x\rangle\langle x|^{X}\otimes\rho^{A}_{x}\otimes\rho^{B}_{x}$ be two tripartite states classical on $X$ system. Let $\varepsilon\in(0,1)$. Then, the hypothesis testing conditional mutual information is defined as:
\begin{align*}
I_{\h}^{\varepsilon}(A;B|X)_{\rho}\coloneqq D_{\h}^{\varepsilon}(\rho^{XAB}\|\rho^{A-X-B}).
\end{align*}
\end{definition}
From Lemma \ref{hrelative}, the following can be seen:
\begin{align}
\label{relationHc}
I_{\h}^{\varepsilon}(A;B|X)_{\rho}\leq\frac{1}{1-\varepsilon}(I(A;B|X)_{\rho}+h_{b}(\varepsilon)).
\end{align}
Notice that for states $\rho^{XAB}\coloneqq\sum_{x}p(x)|x\rangle\langle x|^{X}\otimes\rho^{AB}_{x}$ and $\rho^{A-X-B}\coloneqq\sum_{x}p(x)|x\rangle\langle x|^{X}\otimes\rho^{A}_{x}\otimes\rho^{B}_{x}$, we have $D(\rho^{XAB}\|\rho^{A-X-B})=I(A;B|X)_{\rho}$.

\begin{definition}[Max-relative entropy \cite{Nilanjana-2009}]
\label{mre}
For quantum states $\rho$ and $\sigma$, the max-relative entropy is defined as follows:
\begin{align}
D_{\ma}(\rho\|\sigma)\coloneqq\inf\left\{\lambda\in\mathbbm{R} : \rho\leq 2^{\lambda}\sigma\right\},
\end{align}
where it is well-defined if $\text{supp}(\rho)\subseteq\text{supp}(\sigma)$.
\end{definition}
\begin{lemma}[\cite{Nilanjana-2009}]
The max-relative entropy is monotonically non-increasing with CPTP maps, i.e., for quantum states $\rho, \sigma$ and any CPTP map $\mathcal{E}$, the following holds:
\begin{align*}
D_{\ma}(\mathcal{E}(\rho),\mathcal{E}(\sigma))\leq D_{\ma}(\rho,\sigma).
\end{align*}
\end{lemma}
\begin{definition}[Smooth max-relative entropy \cite{Nilanjana-2009}]
\label{max1}
For a parameter $\epsilon\in (0, 1)$ and quantum states $\rho$ and $\sigma$, the smooth max-relative entropy is defined as:
\begin{align*}
D_{\ma}^{\varepsilon}(\rho\|\sigma)\coloneqq\min_{\substack{\rho^{\prime}\in\cB(\rho)}}D_{\ma}(\rho^{\prime}\|\sigma).
\end{align*}
\end{definition}
From the smooth max-relative entropy, one can define a mutual information-like quantity for a bipartite state $\rho^{AB}$ as follows:
\begin{align}
\label{mquantity1}
D_{\ma}^{\varepsilon}(A;B)_{\rho}\coloneqq D_{\ma}^{\varepsilon}(\rho^{AB}\|\rho^{A}\otimes\rho^{B})=\min_{\substack{\rho^{\prime}\in\cB(\rho)}}D_{\ma}(\rho'^{AB}\|\rho^{A}\otimes\rho^{B}).
\end{align}
\begin{lemma}
\label{mrelative}
For quantum states $\rho$ and $\sigma$ and a parameter $\varepsilon\in(0,1)$, the following indicates the relation between the smooth max-relative entropy and quantum relative entropy.
\begin{align*}
D_{\ma}^{\sqrt{2\varepsilon}}(\rho\|\sigma)\leq\frac{1}{1-\varepsilon}(D(\rho\|\sigma)+h_{b}(\varepsilon)),
\end{align*}
where $h_{b}(\varepsilon)\coloneqq-\varepsilon\log\varepsilon-(1-\varepsilon)\log(1-\varepsilon)$ is the binary entropy function. For a restricted set of values of $\varepsilon$, namely, $\varepsilon\in(0,\frac{1}{\sqrt{2}}]$, we also have the following:
\begin{align*}
D_{\ma}^{\varepsilon}(\rho\|\sigma)\geq D(\rho\|\sigma).
\end{align*}
\begin{IEEEproof}
The proof of the first inequality, the upper bound on the smooth max-relative entropy follows by a straightforward manipulation of Proposition 4.1 in \cite{2012arXiv1211.3141D} and Lemma \ref{hrelative} above. To prove the second inequality, note the following second-order asymptotic of the smooth max-relative entropy \cite{6574274}, \cite{li2014}
\begin{align*}
D_{\ma}^{\varepsilon}(\rho^{\otimes n}\|\sigma^{\otimes n})=nD(\rho\|\sigma)-\sqrt{nV(\rho\|\sigma)}\Phi^{-1}(\varepsilon^{2})+O(\log n),
\end{align*}
where $V(\rho\|\sigma)\coloneqq \Tr\rho(\log\rho-\log\sigma)^{2}\geq0$ is the quantum information variance, $\Phi^{-1}(.)$ is the inverse of the cumulative distribution of the standard normal random variable and $O(\log n)$ lies between a constant and $2\log n$. The proof follows by inserting $n=1$ in the second-order asymptotic and restricting the values of $\varepsilon$ such that the second term on the right-hand side is positive. It can be easily verified that the inverse function becomes negative when its argument is less that $1/2$, therefore we will have $\varepsilon^{2}\leq\frac{1}{2}$. This concludes the proof.  
\end{IEEEproof}
\end{lemma}
\begin{definition}[\cite{ANR2017}] 
\label{max2}
For a bipartite state $\rho^{AB}$ and a parameter $\varepsilon\in(0, 1)$, a mutual information-like quantity can be defined as follows: 
\begin{align*}
\widetilde{I}_{\ma}^{\varepsilon}(A;B)_{\rho}&\coloneqq\inf_{\substack{\rho'^{AB}\in\cB(\rho^{AB})}}D_{\ma}(\rho'^{AB}\|\rho'^{A}\otimes\rho^{B}).
\end{align*}
\end{definition}
The following lemmas relate the aforementioned mutual information-like quantity and the quantity defined in (\ref{mquantity1}).
\begin{lemma}[\cite{ANR2017}]
\label{upper3}
For a bipartite state $\rho^{AB}$ and a parameter $\varepsilon\in(0, 1)$, the following relation holds:
\begin{align*}
\widetilde{I}_{\ma}^{2\varepsilon}(A; B)_{\rho}\leq D_{\ma}^{\varepsilon}(A; B)_{\rho}+\log_{2}\left(\frac{3}{\varepsilon^{2}}\right).
\end{align*}
\end{lemma}
\begin{lemma}
\label{upper1}
For a bipartite state $\rho^{AB}$ and a parameter $\varepsilon\in(0, 1)$, the following relation holds:
\begin{align*}
D_{\ma}^{\varepsilon}(A;B)_{\rho}\leq\widetilde{I}_{\ma}^{\varepsilon}(A;B)_{\rho}.
\end{align*}
\begin{IEEEproof}
The proof is given in the appendix.
\end{IEEEproof}
\end{lemma}
The following mutual information-like quantities can be considered as conditional forms of the quantities given in Definition \ref{max1} and Definition \ref{max2}. 
\begin{definition}
\label{maxmutual}
Let $\rho^{XAB}\coloneqq\sum_{x}p(x)|x\rangle\langle x|^{X}\otimes\rho^{AB}_{x}$ and $\rho^{A-X-B}\coloneqq\sum_{x}p(x)|x\rangle\langle x|^{X}\otimes\rho^{A}_{x}\otimes\rho^{B}_{x}$ be quantum states classical on $X$ and $\varepsilon\in(0,1)$, then
\begin{align*}
D_{\ma}^{\varepsilon}(A;B|X)_{\rho}\coloneqq D_{\ma}^{\varepsilon}(\rho^{XAB}\|\rho^{A-X-B})_{\rho}.
\end{align*}
From Lemma \ref{mrelative}, the following relations can be seen:
\begin{align}
D_{\ma}^{\sqrt{2\varepsilon}}(A;B|X)_{\rho}\leq\frac{1}{1-\varepsilon}(I(A;B|X)+h_{b}(\varepsilon)),
\end{align}
and for $\varepsilon\in(0,\frac{1}{\sqrt{2}}]$, we have
\begin{align}
\label{relationMc}
D^{\varepsilon}_{\ma}(A;B|X)_{\rho}\geq I(A;B|X)_{\rho}.
\end{align}
We define another mutual information-like quantity similar to the one given by Definition \ref{maxmutual}.
\begin{definition}
Let $\rho^{XAB}\coloneqq\sum_{x}p(x)|x\rangle\langle x|^{X}\otimes\rho^{AB}_{x}$ and $\rho^{A-X-B}\coloneqq\sum_{x}p(x)|x\rangle\langle x|^{X}\otimes\rho^{A}_{x}\otimes\rho^{B}_{x}$ be quantum states classical on $X$ and $\varepsilon\in(0,1)$, then
\begin{align*}
\widetilde{I}_{\ma}^{\varepsilon}(A;B|X)_{\rho}\coloneqq \min_{\substack{\rho'\in\cB(\rho)}}(\rho'^{XAB}\|\sum_{x}p'(x)|x\rangle\langle x|^{X}\otimes\rho'^{A}_{x}\otimes\rho^{B}_{x}),
\end{align*}
where $\Tr_{B}\rho'^{XAB}=\sum_{x}p'(x)|x\rangle\langle x|^{X}\otimes\rho'^{A}_{x}$.
\end{definition}
\end{definition}

\begin{remark}
\label{pinch}
In the definition above, it is implied that the minimization is in fact being performed over states which are classical on $X$ subsystem, leading to the conclusion that the optimal state attaining the minimum is classical on $X$. Lemma 6.6 in \cite{Marcobook} studied two important entropic quantities, namely smooth conditional min- and max-entropies, and concluded that smoothing respects the structure of the state $\rho^{XAB}$, meaning that the optimal state $\rho'^{XAB}\in\cB(\rho^{XAB})$ will be classical on $X$ subsystem. Here we make an argument showing that our definition is indeed a legitimate definition. Let $\bar{\rho}^{XAB}\in\cB(\rho^{XAB})$ be the state attaining the minimum in the quantity $D_{\ma}^{\varepsilon}\big(\bar{\rho}^{XAB}\|\bar{\rho}^{XA}\otimes(\sum_{x}\ketbra{x}\otimes\rho^{B}_{x})\big)$. Consider the pinching map $\mathscr{P}^{X}(.)=\sum_{x}\ketbra{x}.\ketbra{x}$. Let $\rho'^{XAB}=\mathscr{P}^{X}(\bar{\rho}^{XAB})$. Note that the pinching map does not affect $\rho^{XAB}$, and since such maps are CPTP and unital, from the monotonicity of the purified distance and also smooth max-relative entropy, we will have $\rho'^{XAB}\in\cB(\rho^{XAB})$ and $\widetilde{I}_{\ma}^{\varepsilon}(A;B|X)_{\rho'XAB}\leq D_{\ma}^{\varepsilon}\big(\bar{\rho}^{XAB}\|\bar{\rho}^{XA}\otimes(\sum_{x}\ketbra{x}\otimes\rho^{B}_{x})\big)$. This concludes that in the minimization of $D_{\ma}^{\varepsilon}\big(\bar{\rho}^{XAB}\|\bar{\rho}^{XA}\otimes(\sum_{x}\ketbra{x}\otimes\rho^{B}_{x})\big)$, one can consider states that are classical on $X$ subsystem.
\end{remark}
\begin{lemma}
\label{upper2}
For quantum states $\rho^{XAB}=\sum_{x}p(x)\ketbra{x}\otimes\rho_{x}^{AB}$ and $\rho^{A-X-B}\coloneqq\sum_{x}p(x)|x\rangle\langle x|^{X}\otimes\rho^{A}_{x}\otimes\rho^{B}_{x}$ and a parameter $\varepsilon\in(0,1)$, we have:
\begin{align*}
\widetilde{I}_{\ma}^{2\varepsilon}(A;B|X)_{\rho}\leq D_{\ma}^{\varepsilon}(A;B|X)_{\rho}+\log(\frac{1}{1-\sqrt{1-\varepsilon^{2}}}+1).
\end{align*}
\begin{IEEEproof}
The proof is relegated to the appendix.
\end{IEEEproof}
\end{lemma}

\begin{lemma}\footnote{Note that for our purposes in this paper, the upper bound given by Lemma \ref{upper2} is enough; We prove this lemma further for sake of completeness of our study.}
\label{lower2}
For quantum states $\rho^{XAB}=\sum_{x}p(x)\ketbra{x}\otimes\rho_{x}^{AB}$ and $\rho^{A-X-B}\coloneqq\sum_{x}p(x)|x\rangle\langle x|^{X}\otimes\rho^{A}_{x}\otimes\rho^{B}_{x}$ and a parameter $\varepsilon\in(0,1)$, the following stands:
\begin{align*}
D_{\ma}^{\varepsilon}(A;B|X)_{\rho}\leq\widetilde{I}_{\ma}^{\varepsilon}(A;B|X)_{\rho}.
\end{align*} 
\begin{IEEEproof}
The proof is provided in the appendix.
\end{IEEEproof}
\end{lemma}
\begin{lemma}[\cite{6574274},\cite{li2014}]
\label{asym3}
For quantum states $\rho,\sigma$ and a parameter $\varepsilon\in(0,1)$, we have:
\begin{align*}
\lim_{n\rightarrow\infty}\frac{1}{n}D_{\ma}^{\varepsilon}(\rho^{\otimes n}\|\sigma^{\otimes n})&=D(\rho\|\sigma),\\
\lim_{n\rightarrow\infty}\frac{1}{n}D_{\h}^{\varepsilon}(\rho^{\otimes n}\|\sigma^{\otimes n})&=D(\rho\|\sigma).
\end{align*}
The followings are straightforward consequences of Lemma \ref{asym3}. For quantum state $\rho^{AB}$ and a parameter $\varepsilon\in(0,1)$, we have: 
\begin{align*}
\lim_{n\rightarrow\infty}\frac{1}{n}D_{\ma}^{\varepsilon}(A;B)_{\rho^{\otimes n}}=D(\rho^{AB}\|\rho^{A}\otimes\rho^{B})=I(A;B)_{\rho},
\end{align*}
\begin{align*}
\lim_{n\rightarrow\infty}\frac{1}{n}I_{\h}^{\varepsilon}(A;B)_{\rho^{\otimes n}}=D(\rho^{AB}\|\rho^{A}\otimes\rho^{B})=I(A;B)_{\rho}.
\end{align*}
And for the quantum states $\rho^{XAB}\coloneqq\sum_{x}p(x)|x\rangle\langle x|^{X}\otimes\rho^{AB}_{x}$ and $\rho^{A-X-B}\coloneqq\sum_{x}p(x)|x\rangle\langle x|^{X}\otimes\rho^{A}_{x}\otimes\rho^{B}_{x}$, we have:
\begin{align}
\label{asymmm}
\lim_{n\rightarrow\infty}\frac{1}{n}D_{\ma}^{\varepsilon}(A;B|X)_{\rho^{\otimes n}}&=D(\rho^{XAB}\|\rho^{A-X-B})=I(A;B|X)_{\rho},\\ \nonumber
\lim_{n\rightarrow\infty}\frac{1}{n}I_{\h}^{\varepsilon}(A;B|X)_{\rho^{\otimes n}}&=D(\rho^{XAB}\|\rho^{A-X-B})=I(A;B|X)_{\rho}.
\end{align}
\end{lemma}
\begin{lemma}
\label{asym4}
For quantum states $\rho^{XAB}=\sum_{x}p(x)\ketbra{x}\otimes\rho_{x}^{AB}$ and $\rho^{A-X-B}\coloneqq\sum_{x}p(x)|x\rangle\langle x|^{X}\otimes\rho^{A}_{x}\otimes\rho^{B}_{x}$ and a parameter $\varepsilon\in(0,1)$, it holds that:
\begin{align*}
\lim_{n\rightarrow\infty}\frac{1}{n}\widetilde{I}_{\ma}^{\varepsilon}(A;B|X)_{\rho^{\otimes n}}=I(A;B|X)_{\rho}.
\end{align*}
\begin{IEEEproof}
The proof follows from Lemma \ref{upper2} and \ref{lower2} as well as the fact given by (\ref{asymmm}). 
\end{IEEEproof}
\end{lemma}

The following lemma comes in handy in the proof of the conditional convex-split lemma.
\begin{lemma}
\label{convexproof}
For an ensemble of cq states $\{\rho^{XA}_{1},...,\rho^{XA}_{n}\}$ and a probability mass function  $\{p(i)\}_{i=1}^{n}$, let $\rho^{XA}=\sum_{i}p(i)\rho^{XA}_{i}$ be the average state. Then for a state $\theta^{XA}$ we have the following equality:
\begin{align*}
D(\rho^{XA}||\theta^{XA})=\sum_{i=1}^{n}p(i)\left(D(\rho^{XA}_{i}\|\theta^{XA})-D(\rho^{XA}_{i}||\rho^{XA})\right).
\end{align*}
\begin{IEEEproof}
Proof is presented in the appendix.
\end{IEEEproof}
\end{lemma}
\begin{lemma}[Conditional convex-split lemma]
\label{condconex}
Consider the cq state $\rho^{XAB}\coloneqq\sum_{x}p(x)|x\rangle\langle x|^{X}\otimes\rho^{AB}_{x}$,  define $\sum_{x}p(x)|x\rangle\langle x|^{X}\otimes\rho^{A}_{x}\otimes\sigma^{B}_{x}$ such that supp$(\rho_{x}^{B})\subseteq \text{supp}(\sigma_{x}^{B})$ for all $x$. Let 
$k\coloneqq D_{\ma}(\rho^{XAB},\sum_{x}p(x)|x\rangle\langle x|^{X}\otimes\rho^{A}_{x}\otimes\sigma^{B}_{x})$. Define the following state:
\begin{align*}
\tau^{XAB_{1}...B_{n}}\coloneqq\sum_{x}p(x)|x\rangle\langle x|^{X}\otimes\big(\frac{1}{n}\sum_{j=1}^{n}\rho^{AB_{j}}_{x}\otimes\sigma^{B_{1}}_{x}\otimes...\otimes\sigma^{B_{j-1}}_{x}\sigma^{B_{j+1}}_{x}\otimes\sigma^{B_{n}}_{x}\big),
\end{align*}
on $n+2$ systems $X,A,B_{1},...,B_{n}$, where for $\forall j\in[1:n]$ and $x\in \text{supp}(p(x)):\rho^{AB_{j}}_{x}=\rho^{AB}_{x}$ and $\sigma^{B_{j}}_{x}=\sigma^{B}_{x}$. We have the following:
\begin{align*}
D\big(\tau^{XAB_{1}...B_{n}}\big|\big|\sum_{x}p(x)|x\rangle\langle x|^{X}\otimes\rho^{A}_{x}\otimes\sigma^{B_{1}}_{x}\otimes...\otimes\sigma^{B_{n}}_{x}\big)\leq\log(1+\frac{2^{k}}{n}).
\end{align*} 
In particular, for some $\delta\in(0,1)$ and $n=\ceil{\frac{2^{k}}{\delta^{2}}}$ the following holds:
\begin{align*}
P(\tau^{XAB_{1}...B_{n}},\sum_{x}p(x)|x\rangle\langle x|^{X}\otimes\rho^{A}_{x}\otimes\sigma^{B_{1}}_{x}\otimes...\otimes\sigma^{B_{n}}_{x})\leq\delta.
\end{align*}
\begin{IEEEproof}
The proof is presented in the appendix.
\end{IEEEproof}
\end{lemma}
We saw that the conditional-max relative entropy naturally appeared in the conditional convex-split lemma. The importance of the smooth entropies have been widely recognized. In the following, we present a variation of the conditional convex-split lemma which involves smooth conditional max-relative entropy.
\begin{corollary}
\label{cofcondconvex}
Fix a $\varepsilon>0$. Let $\rho^{XAB}=\sum_{x}p(x)\ketbra{x}^{X}\otimes\rho^{AB}_{x}$ and $\sum_{x}p(x)\ketbra{x}^{X}\otimes\rho^{A}_{x}\otimes\sigma^{B}_{x}$ be quantum states such that $\text{supp}(\rho^{B}_{x})\subseteq \text{supp}(\sigma^{B}_{x})$ for all $x$. Define $k\coloneqq \min_{\substack{\rho'\in\cB(\rho)}}D_{\ma}(\rho'^{XAB}\|\sum_{x}p'(x)|x\rangle\langle x|^{X}\otimes\rho'^{A}_{x}\otimes\sigma^{B}_{x})$ where the optimization takes place over states classical on $X$. Further define the following state
\begin{align*}
\tau^{XAB_{1}...B_{n}}\coloneqq\sum_{x}p(x)|x\rangle\langle x|^{X}\otimes\big(\frac{1}{n}\sum_{j=1}^{n}\rho^{AB_{j}}_{x}\otimes\sigma^{B_{1}}_{x}\otimes...\otimes\sigma^{B_{j-1}}_{x}\otimes\sigma^{B_{j+1}}_{x}\otimes\sigma^{B_{n}}_{x}\big),
\end{align*}
on $n+2$ systems $X,A,B_{1},...,B_{n}$, where $\forall j\in[1:n]$ and $x\in \text{supp}(p(x)):\rho^{AB_{j}}_{x}=\rho^{AB}_{x}$ and $\sigma^{B_{j}}_{x}=\sigma^{B}_{x}$. For $\delta\in(0,1)$ and $n=\ceil{\frac{2^{k}}{\delta^{2}}}$, the following holds true:
\begin{align*}
P(\tau^{XAB_{1}...B_{n}},\sum_{x}p(x)\ketbra{x}^{X}\otimes\rho^{A}_{x}\otimes\sigma^{B_{1}}_{x}\otimes...\otimes\sigma^{B_{n}}_{x})\leq 2\varepsilon+\delta.
\end{align*}
\begin{IEEEproof}
Proof is presented in the appendix.
\end{IEEEproof}
\end{corollary} 

\begin{lemma}[Hayashi-Nagaoka operator inequality \cite{Hayashi-Nagaoka-2003}]
\label{HN}
Let $T, S\in\mathcal{P}(\mathcal{H}^{A})$ such that $(\mathbbm{1}-S)\in\mathcal{P}(\mathcal{H}^{A})$. Then for all constants $c>0$, the following inequality holds:
\begin{align*}
\mathbbm{1}-(S+T)^{-\frac{1}{2}}S&(S+T)^{-\frac{1}{2}}\\
&\leq(1+c)(\mathbbm{1}-S)+(2+c+c^{-1})T.
\end{align*}
\end{lemma}

 \section{Information-Processing Task, Code Definition and Main Results}
 \begin{figure}
\begin{center}
\includegraphics[width=0.7\textwidth]{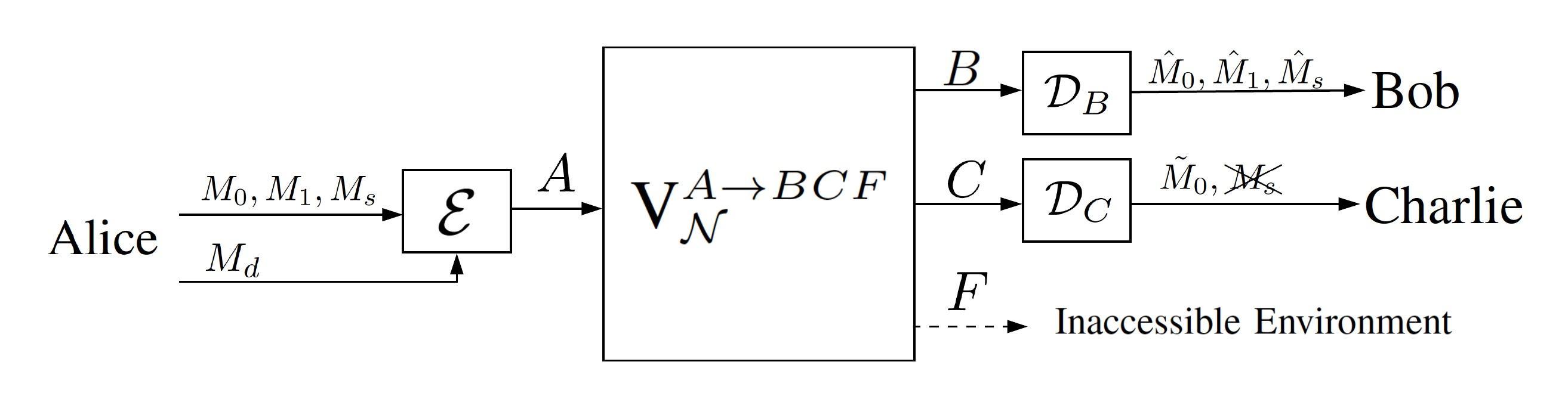}      
\caption{Single-serving quantum broadcast channel with isometric extension $V^{A\rightarrow BCF}_{\mathcal{N}}$. Alice attempts to transmit a common message $M_{0}$ to Bob and Charlie and a private message $M_{1}$ and a confidential message $M_{s}$ to Bob only such that the confidential message must be kept secret from Charlie. The dummy randomness used by Alice for encryption is modeled by a message $M_{d}$.}
\label{BC}
\end{center}
\end{figure}
Consider the quantum broadcast communication system model depicted in Fig. \ref{BC}. A quantum broadcast channel $\mathcal{N}^{A\rightarrow BC}$ with isometric extension $\mbox{\normalsize V}^{A\rightarrow BCF}_{\mathcal{N}}$ connects a sender in possession of $A$ system (Alice) to two receivers, a primary receiver (Bob) in possession of $B$ and a third-party receiver (Charlie) possessing $C$ system and the communication is surrounded by an inaccessible environment modeled as $F$ system. Alice attempts to send three messages simultaneously: a common message $M_{0}$ that is supposed to be decoded by both Bob and Charlie, a private message $M_{1}$ that is intended to Bob with no secrecy requirement imposed upon it and a confidential message $M_{s}$ exclusive to Bob that must not be leaked to Charlie. The obfuscation of the confidential message is done by virtue of stochastic encoding, i.e., introducing randomness into codewords in the encoding process. It is convenient to represent this randomness as the realization of a discrete memoryless source which is independent of the channel and the messages to be transmitted. We find it even more useful to think of the so-called randomness as a \textit{dummy message} $M_{d}$ taking its values according to some distribution\footnote{It will be seen that the difference between the private and dummy messages is whether Bob wants to decoded it or not.}.

The encoder encodes the message triple $(M_{0},M_{1},M_{s})$ as well as the dummy message $M_{d}$ into a quantum codeword $A$ and transmits it over the channel. Upon receiving $B$ and $C$ systems, Bob finds the estimates $\hat{M}_{0},\hat{M}_{1},\hat{M}_{s}$ of the common, private and confidential messages, respectively, while Charlie finds the estimate $\tilde{M}_{0}$ of the common message. To ensure reliability and security, a tradeoff arises between the rates of the messages. We study the one-shot limit on this tradeoff.

\begin{definition} A $(2^{R_{0}},2^{R_{1}},2^{R_{s}})$ one-shot code $\mathcal{C}$ for the quantum broadcast channel $\mathcal{N}^{A\rightarrow BC}$ consists of 
\begin{itemize}
\item Three message sets $[1 : 2^{R_{0}}]$, $[1 : 2^{R_{1}}]$ and $[1 : 2^{R_{s}}]$ (common, private and confidential, respectively),
\item A source of local randomness $[1 : 2^{R_{d}}]$,
\item An encoding operator $\mathcal{E}: M_{0}\times M_{1}\times M_{s}\times M_{d}\rightarrow A$, which maps a message triple $(m_{0},m_{1},m_{s})\in [1 : 2^{R_{0}}]\times [1 : 2^{R_{1}}]\times[1 : 2^{R_{s}}]$ and a realization of the local source of randomness $m_{d}\in[1 : 2^{R_{d}}]$ to a codeword $\rho^{A}$,
\item A decoding POVM $\mathcal{D}_{B}: B\rightarrow  (M_{0}\times M_{1}\times M_{s}) \cup \{?\}$,  which assigns an estimate $(\hat{m}_{0},\hat{m}_{1},\hat{m}_{s})\in [1 : 2^{R_{0}}]\times [1 : 2^{R_{1}}]\times[1 : 2^{R_{s}}]$ or an error message $\{?\}$ to each received state $\rho^{B}$,
\item A decoding POVM $\mathcal{D}_{C}: C\rightarrow M_{0} \cup \{?\}$ that assigns an estimate $\tilde{m}_{0}\in[1 : 2^{R_{0}}]$ or an error message $\{?\}$ to each received state $\rho^{C}$.
\end{itemize}
\end{definition}

The $(2^{R_{0}},2^{R_{1}},2^{R_{s}})$ one-shot code is assumed to be known by all parties ahead of time; Likewise, the statistics of the source of randomness are assumed known to all parties, however, its realizations used in the encoding process are only accessible by Alice. Note that we have included the source of randomness in the definition of the code to imply that it can be optimized over as part of the code design. However, we do not consider the effect of non-uniform randomness in our analysis \cite{Bloch-Kliewer} and throughout we assume that the dummy message $M_{d}$ is uniformly distributed over $[1:2^{R_{d}}]$.  We further assume that the message triple $(M_{0},M_{1},M_{d})$ is uniformly distributed over $[1 : 2^{R_{0}}]\times [1 : 2^{R_{1}}]\times[1 : 2^{R_{s}}]$ so that the rates of the common, private and confidential messages are $H(M_{0})=R_{0}, H(M_{1})=R_{1}$ and $H(M_{s})=R_{s}$, respectively. The reliability performance of the code $\mathcal{C}$ is measured by its average probability of error defined as follows:
\begin{align}
\label{reliability}
P_{\text{error}}^{1}\coloneqq\text{Pr}\{(\hat{M}_{0},\hat{M}_{1},\hat{M}_{s})\neq(M_{0},M_{1},M_{s})\;\text{or}\;\tilde{M}_{0}\neq M_{0}\},
\end{align}
while its secrecy level, i.e., an indication of Charlie's ignorance about the confidential message, is measured in terms of the trace distance between Charlie's received state and some constant state as follows:
\begin{align}
\label{privacy}
\forall m_{0}:\quad P_{\text{secrecy}}^{1}(m_{0})\coloneqq\frac{1}{2^{R_{s}}}\sum_{m_{s}}\frac{1}{2}\|\rho^{C}_{m_{0},m_{s}}-\sigma^{C}_{m_{0}}\|_{1}.
\end{align} 
Note that the secrecy requirement indicates Charlie's ignorance about the confidential message $m_{s}$ on average conditioned on the fact that he has decoded the common message $m_{0}$ correctly.

A rate quadruple $(R_{0},R_{1},R_{s},R_{d})$ is said to be $\varepsilon$-achievable if there exist a one-shot code $\mathcal{C}$ satisfying the following conditions:
\begin{align}
\label{error}
P_{\text{error}}^{1}&\leq\varepsilon,\\
\label{secrecy}
\forall m_{s}:P_{\text{secrecy}}^{1}(m_{0})&\leq\varepsilon, 
\end{align}
where $\varepsilon\in(0,1)$ characterizes both the reliability and secrecy of the code. Then the $\varepsilon$-achievable rate region $\mathcal{R}^{\varepsilon}(\mathcal{N})$ is defined to consists of the closure of the set of all $\epsilon$-achievable rate quadruples. In this paper, our main goal is to find the optimal rate region $\mathcal{R}^{\varepsilon}(\mathcal{N})$ by establishing achievability and converse bounds.

The following theorem presents our achievability bound on $\mathcal{R}^{\epsilon}(\mathcal{N})$.
\begin{theorem}[Achievability Region]
\label{achievability}
Fix $\varepsilon', \varepsilon'',\delta_{1},\delta_{2},\delta_{3}$ and $\eta$ such that $0<3\varepsilon'+2\sqrt{\varepsilon'}<1,0<\delta_{1},\delta_{2},\delta_{3}<\varepsilon'$, $0<\varepsilon''<\sqrt{2}-1,$ $0<\eta<\varepsilon''^{2}$. Consider a quantum broadcast channel $\mathcal{N}^{A\rightarrow BC}$. Let the random variables $U,V$ and $X$ be distributed according to a distribution $p(u,v,x)$ which factorizes as $p(u,v,x)=p(u,v)p(x|v)$ and define cq state $\rho^{UVXA}=\sum_{u,v,x}p(u,v,x)\ketbra{u}^{U}\otimes\ketbra{v}^{V}\otimes\ketbra{x}^{X}\otimes\rho_{x}^{A}$. Let $\mathcal{R}^{(\text{in})}(\rho)$ be the set of those quadruples $(R_{0},R_{1},R_{s},R_{d})$ satisfying the following conditions on $\rho^{UVXBC}=\mathcal{N}^{A\rightarrow BC}(\rho^{UVXA})$:
\begin{align}
\label{common}
R_{0}&\leq\text{min}\big[I^{\varepsilon'-\delta_{1}}_{\h}(U;B)_{\rho}-\log_{2}(\frac{4\varepsilon'}{\delta_{1}^{2}}),I_{\h}^{\varepsilon'-\delta_{2}}(U;C)_{\rho}-\log_{2}(\frac{4\varepsilon'}{\delta_{2}^{2}})\big], \\
\label{all}
R_{0}+R_{1}+R_{s}&\leq I_{\h}^{\varepsilon'-\delta_{3}}(V;B|U)_{\rho}-\log_{2}(\frac{4\varepsilon'}{\delta_{3}^{2}})+\text{min}\big[I^{\varepsilon'-\delta_{1}}_{\h}(U;B)_{\rho}-\log_{2}(\frac{4\varepsilon'}{\delta_{1}^{2}}),I_{\h}^{\varepsilon'-\delta_{2}}(U;C)_{\rho}-\log_{2}(\frac{4\varepsilon'}{\delta_{2}^{2}})\big], \\
\label{confi}
R_{s}&\leq I_{\h}^{\varepsilon'-\delta_{3}}(V;B|U)_{\rho}-\widetilde{I}_{\ma}^{\varepsilon''}(V;C|U)_{\rho}-\log_{2}(\frac{4\varepsilon'}{\delta_{3}^{2}})-2\log_{2}(\frac{1}{\eta}), \\
\label{convex2}
R_{1}+R_{d}&\geq\widetilde{I}_{\ma}^{\varepsilon''}(V;C|U)_{\rho}+\widetilde{I}_{\ma}^{\varepsilon''}(X;C|V)_{\rho}+4\log_{2}(\frac{1}{\eta}),\\
\label{convex1}
R_{d}&\geq \widetilde{I}_{\ma}^{\varepsilon''}(X;C|V)_{\rho}+2\log_{2}(\frac{1}{\eta}).
\end{align}
Let $\varepsilon\coloneqq\max\{\sqrt[4]{\varepsilon'},\sqrt[4]{\varepsilon''}\}$. Then $\bigcup\mathcal{R}^{(\text{in})}(\rho)\subseteq\mathcal{R}^{\varepsilon}(\mathcal{N})$ where the union is over all $\rho^{UVXBC}$ arising from the channel.
\end{theorem}
\begin{theorem}[Converse Region]
\label{converse1}
Fix $\varepsilon\in(0,\frac{1}{4}]$. Let the random variables $U,V$ and $X$ be distributed according to a distribution $p(u,v,x)$ which factorizes as $p(u,v,x)=p(u,v)p(x|v)$ and define cq state $\rho^{UVXA}=\sum_{u,v,x}p(u,v,x)\ketbra{u}^{U}\otimes\ketbra{v}^{V}\otimes\ketbra{x}^{X}\otimes\rho_{x}^{A}$. Let the state $\rho^{UVXBC}$ be the result of the action of the quantum broadcast channel $\mathcal{N}^{A\rightarrow BC}$ on the state $\rho^{UVXA}$. Let $\mathcal{R}^{(\text{co})}(\rho)$ be the set of those quadruples $(R_{0},R_{1},R_{s},R_{d})$ satisfying the following conditions:
\begin{align}
\label{commonc}
R_{0}&\leq\text{min}\big[I^{\varepsilon}_{\h}(U;B)_{\rho},I_{\h}^{\varepsilon}(U;C)_{\rho}\big], \\
\label{allc}
R_{0}+R_{1}+R_{s}&\leq I_{\h}^{\varepsilon}(V;B|U)_{\rho}+\text{min}\big[I^{\varepsilon}_{\h}(U;B)_{\rho},I_{\h}^{\varepsilon}(U;C)_{\rho}\big], \\
\label{confic}
R_{s}&\leq I_{\h}^{\varepsilon}(V;B|U)_{\rho}-D_{\ma}^{\sqrt{2\varepsilon}}(V;C|U)_{\rho}, \\
\label{convex2c}
R_{1}+R_{d}&\geq D_{\ma}^{\sqrt{2\varepsilon}}(V;C|U)_{\rho}+D_{\ma}^{\sqrt{2\varepsilon}}(X;C|V)_{\rho},\\
\label{convex1c}
R_{d}&\geq D_{\ma}^{\sqrt{2\varepsilon}}(X;C|V)_{\rho}.
\end{align}
Then $\mathcal{R}^{\varepsilon}(\mathcal{N})\subseteq\bigcup\mathcal{R}^{(\text{co})}(\rho)$ and the union is over all $\rho^{UVXBC}$ arising from the channel.

From the theorems above, the recent result of the current authors on the simultaneous transmission of classical and quantum information can be recovered. The slight difference between the results stems from the fact that in \cite{Farzin.arXiv.2018}, there is a single criterion for the error probability and secrecy while in this work separate criteria are considered.
\end{theorem}
\begin{corollary}[\cite{Farzin.arXiv.2018}]
\label{simulone}
Fix $\varepsilon', \varepsilon'',\delta_{1},\delta_{3}$ and $\eta$ such that $0<3\varepsilon'+2\sqrt{\varepsilon'}<1,0<\delta_{1},\delta_{3}<\varepsilon'$, $0<\varepsilon''<\sqrt{2}-1,$ $0<\eta<\varepsilon''^{2}$. Define $\varepsilon\coloneqq\max\{\sqrt[4]{\varepsilon'},\sqrt[4]{\varepsilon''}\}$. Let $C^{\varepsilon}$ denote the one-shot capacity region of the channel $\mathcal{N}^{A\rightarrow BE}$ for simultaneous transmission of classical and quantum information. For a cq state $\rho^{UVA}$ classical on $U$ and $V$ subsystems, the following achievability bound holds:
\begin{align*}
C^{(in)}\subseteq C^{\varepsilon},
\end{align*}  
where, denoting the one-shot rates of the classical and quantum information by $R_{c}^{1}$ and $R_{q}^{1}$, respectively, $C^{(in)}$ is the union over all states $\rho^{UVBE}$ arising from the channel, of rate pairs $(R_{c}^{1},R_{q}^{1})$ obeying:
\begin{align*}
R_{c}^{1}&\leq I^{\varepsilon'-\delta_{1}}_{\h}(U;B)_{\rho}-\log_{2}(\frac{4\varepsilon'}{\delta_{1}^{2}}), \\
R_{q}^{1}&\leq I_{\h}^{\varepsilon'-\delta_{3}}(V;B|U)_{\rho}-\widetilde{I}_{\ma}^{\varepsilon''}(V;E|U)_{\rho}-\log_{2}(\frac{4\varepsilon'}{\delta_{3}^{2}})-2\log_{2}(\frac{1}{\eta}).
\end{align*}
Redefine $\varepsilon$ as a parameter in $(0,\frac{1}{4}]$. Then the following converse holds:
\begin{align*}
C^{\varepsilon}\subseteq C^{(co)},
\end{align*}
where $C^{(co)}$ is the union over all states $\rho^{UVBE}$ arising from the channel, of rate pairs $(R_{c}^{1},R_{q}^{1})$ obeying
\begin{align*}
R_{c}^{1}&\leq I^{\varepsilon}_{\h}(U;B)_{\rho},\\
R_{q}^{1}&\leq I_{\h}^{\varepsilon}(V;B|U)_{\rho}-D_{\ma}^{\sqrt{2\varepsilon}}(V;E|U)_{\rho}.
\end{align*}
\begin{IEEEproof}
The approach for the simultaneous transmission of classical and quantum information is through finding the limits on the simultaneous transmission of common and confidential messages. From \cite{Dev-private} it is well-known that the rate of the confidential message can be translated into the rate of quantum information. As hinted in the introductory part, when it comes to transmission of quantum information, there is zero-tolerance condition of copying quantum information, therefore the confidential messages must be kept secret from the entire universe but Bob meaning that the output of the channel consists of a system received by Bob and another inaccessible environment $E$ (which includes Charlie's system). From Theorem \ref{achievability} and Theorem \ref{converse1} onward, since there is no concern regarding the rate of the dummy randomness, the last two inequalities in both regions will be trivial. And the achievability part can be seen from (\ref{common}) and (\ref{confi}) and the converse part from (\ref{commonc}) and (\ref{confic}).
\end{IEEEproof}
\end{corollary}

\section{Achievability}
The first part of the direct coding theorem, the reliability of the messages equation (\ref{error}), is an exquisite combination of the classical superposition coding and position-based decoding. It is well known that the superposition coding suggests a layered encoding approach such that each (possibly independent) message is encoded into a different codebook. On the other hand, in the position-based decoding, the messages are encoded into the positions of quantum states such that the position of each state indicates the message that it contains (in a conservative view though, this also happens to be the case in the ordinary channel coding where each message is encoded into a particular row of the codebook matrix). The second part of the direct coding theorem, the secrecy of the confidential message equation (\ref{secrecy}), is handled by a version of the convex-split lemma which relies on superposition of codewords. This should remind us about the channel resolvability via superpositions studied for the classical \cite{Wat-Ooh15} and quantum \cite{Anurag-Hayashi-Warsi.2018} channels. We will prove the channel resolvability via superpositions with virtue of convex-split lemma in this paper.   

In order to establish the achievability of the region put forward by Theorem \ref{achievability}, we first show the achievability of another region and then argue how this region leads to the achievability of the region in Theorem \ref{achievability}.
\begin{lemma}
\label{achievability2}
Fix $\varepsilon', \varepsilon'',\delta_{1},\delta_{2},\delta_{3}$ and $\eta$ such that $0<3\varepsilon'+2\sqrt{\varepsilon'}<1,0<\delta_{1},\delta_{2},\delta_{3}<\varepsilon'$, $0<\varepsilon''<\sqrt{2}-1,$ $0<\eta<\varepsilon''^{2}$ and define $\varepsilon\coloneqq\max\{\sqrt[4]{\varepsilon'},\sqrt[4]{\varepsilon''}\}$. Let the random variables $U,V$ and $X$ be distributed according to a distribution $p(u,v,x)$ which factorizes as $p(u,v,x)=p(u,v)p(x|v)$. We further define cq state $\rho^{UVXA}=\sum_{u,v,x}p(u,v,x)|u\rangle\langle u|^{U}\otimes|v\rangle\langle v|^{V}\otimes|x\rangle\langle x|^{X}\otimes\rho_{x}^{A}$. Let $\mathcal{R}^{*}(\rho)$ be the set of those quadruples $(R_{0},R_{1},R_{s},R_{d})$ satisfying the following conditions on $\rho^{UVXBC}=\mathcal{N}^{A\rightarrow BC}(\rho^{UVXA})$:
\begin{align}
\label{common1}
R_{0}&\leq\text{min}[I^{\varepsilon'-\delta_{1}}_{\h}(U;B)_{\rho}-\log_{2}(\frac{4\varepsilon'}{\delta_{1}^{2}}),I_{\h}^{\varepsilon'-\delta_{2}}(U;C)_{\rho}-\log_{2}(\frac{4\varepsilon'}{\delta_{2}^{2}})], \\
\label{all1}
R_{1}+R_{s}&\leq I_{\h}^{\varepsilon'-\delta_{3}}(V;B|U)_{\rho}-\log_{2}(\frac{4\varepsilon'}{\delta_{3}^{2}}), \\
\label{convex2a}
R_{1}&\geq\widetilde{I}_{\ma}^{\varepsilon''}(V;C|U)_{\rho}+2\log_{2}(\frac{1}{\eta}),\\
\label{convex1a}
R_{d}&\geq \widetilde{I}_{\ma}^{\varepsilon''}(X;C|V)_{\rho}+2\log_{2}(\frac{1}{\eta}),
\end{align}
Then $\bigcup\mathcal{R}^{*}(\rho)\subseteq\mathcal{R}^{\varepsilon}(\mathcal{N})$ and the union is over all $\rho^{UVXBC}$ arising from the channel.
\end{lemma}
\begin{lemma}
\label{fme}
We have $\bigcup_{\rho}\mathcal{R}^{(in)}(\rho)\subseteq\mathcal{R}^{\varepsilon}(\varepsilon)$. 
\begin{IEEEproof}
To prove the lemma we need to show that $\mathcal{R}^{(in)}(\rho)\subseteq\mathcal{R}^{*}(\rho)$. This can be provn in a standard way by Fourier-Motzkin elimination (see for example appendix D of \cite{Gamal:2012:NIT:2181143}); Inequality (\ref{confi}) can be seen from (\ref{all1}) and (\ref{convex2a}). Let $R_{a}\coloneqq R_{1}+R_{d}$ and $R_{b}\coloneqq R_{0}+R_{1}+R_{s}$. Then in the following region, 
\begin{align}
\label{sum1}
R_{b}&\coloneqq R_{0}+R_{1}+R_{s},\\
\label{suma1}
R_{a}&\coloneqq R_{1}+R_{d},\\
\label{common12}
R_{0}&\leq\text{min}[I^{\varepsilon'-\delta_{1}}_{\h}(U;B)_{\rho}-\log_{2}(\frac{4\varepsilon'}{\delta_{1}^{2}}),I_{\h}^{\varepsilon'-\delta_{2}}(U;C)_{\rho}-\log_{2}(\frac{4\varepsilon'}{\delta_{2}^{2}})], \\
\label{all12}
R_{1}+R_{s}&\leq I_{\h}^{\varepsilon'-\delta_{3}}(V;B|U)_{\rho}-\log_{2}(\frac{4\varepsilon'}{\delta_{3}^{2}}), \\
\label{confi2}
R_{s}&\leq I_{\h}^{\varepsilon'-\delta_{3}}(V;B|U)_{\rho}-\widetilde{I}_{\ma}^{\varepsilon''}(V;C|U)_{\rho}-\log_{2}(\frac{4\varepsilon'}{\delta_{3}^{2}})-2\log_{2}(\frac{1}{\eta}), \\
\label{convex2a2}
R_{1}&\geq\widetilde{I}_{\ma}^{\varepsilon''}(V;C|U)_{\rho}+2\log_{2}(\frac{1}{\eta}),\\
\label{convex1a2}
R_{d}&\geq \widetilde{I}_{\ma}^{\varepsilon''}(X;C|V)_{\rho}+2\log_{2}(\frac{1}{\eta}),
\end{align}
one can simply remove (\ref{convex2a2}) by considering (\ref{suma1}) and (\ref{convex1a2}). Likewise, inequality (\ref{all12}) can be removed from (\ref{sum1}) and (\ref{common12}). This leads to the region given by Theorem \ref{achievability}\footnote{Note that the Fourier-Motzkin elimination can also lead to other regions. The region we derived is in accord with the definition of the problem.}.

Alternatively, Lemma \ref{fme} can be shown similar to Lemma 19 of \cite{Wat-Ooh15} using the following argument. From the definition of the problem, if a quadruple $(R_{0}+r_{0},R_{1}-r_{0}-r_{s}+r_{d},R_{s}+r_{s},R_{d}-r_{d})\in\mathcal{R}^{\varepsilon}(\mathcal{N})$ for some $r_{0},r_{s},r_{d}\geq0$, then $(R_{0},R_{1},R_{s},R_{d})\in\mathcal{R}^{\varepsilon}(\mathcal{N})$ as well. Then one can find suitable values of $(r_{0},r_{s},r_{d})$ satisfying the conditions along the same lines as Lemma 19 of \cite{Wat-Ooh15}.
\end{IEEEproof}
\end{lemma}

We begin the proof of Lemma \ref{achievability2} with a sketch of achievability. The coding scheme uses layered encoding and position-based decoding.  Fix $p(u,v,x)$. The classical states associated to the random variables $U,V$ and $X$ will compose the codebooks, which in this case by codebook we mean tensor product states shared among parties. This should resemble the role random variables play in constructing the codebooks in the classical case and superposition of codebooks will be replaced by the superposition of the shared states. The coding consists of three layers, the first layer contains tensor products of $2^{R_{0}}$ copies of $\rho^{U}$. These states will accommodate the common message. Conditioned on each $\rho^{U}$, in the second layer there are $2^{R_{s}+R_{1}}$ copies of $\rho^{V}$. The confidential and private messages (as well as perhaps part of the dummy message) are encoded in this layer. Finally, dummy message is encoded in the third layer into $2^{R^{d}}$ copies of $\rho^{X}$ depending on those in the second layer. Both Bob and Charlie first decode the information in the first layer, i.e., the common message, and then Bob uses the extracted index to find the private and confidential messages. The position-based encoding scheme obviously uses shared randomness ahead of time. After achieving the capacity results, we should derandomize the code by fixing the classical systems and obtaining a protocol that does not rely on shared randomness.  

We now provide the details of the achievability proof. Let $\varepsilon', \varepsilon'',\delta_{1},\delta_{2},\delta_{3}$ and $\eta$ be such that $0<3\varepsilon'+2\sqrt{\varepsilon'}<1,0<\delta_{1},\delta_{2},\delta_{3}<\varepsilon'$, $0<\varepsilon''<\sqrt{2}-1,$ $0<\eta<\varepsilon''^{2}$ and $\varepsilon\coloneqq\max\{\sqrt[4]{\varepsilon'},\sqrt[4]{\varepsilon''}\}$.
\vspace*{5px}

\textbf{Codebook generation:} Fix a pmf $p(u,v,x)=p(u,v)p(x|v)$. Alice, Bob and Charlie share $2^{R_{0}}$ copies of the classical state $\rho^{U^{A}U^{B}U^{C}}\coloneqq\sum_{u}p(u)|u\rangle\langle u|^{U^{A}}\otimes|u\rangle\langle u|^{U^{B}}\otimes|u\rangle\langle u|^{U^{C}}$ as follows:
\begin{align*}
(\rho^{U^{A}U^{B}U^{C}})^{\otimes 2^{R_{0}}}=\rho^{U^{A}_{1}U^{B}_{1}U^{C}_{1}}\otimes...\otimes\rho^{U^{A}_{2^{R_{0}}}U^{B}_{2^{R_{0}}}U^{C}_{2^{R_{0}}}},
\end{align*}
where Alice possesses $U^{A}$ systems, Bob $U^{B}$  systems and Charlie has $U^{C}$ systems (the superscripts should not be confused with the input $A$ or output systems $B$ and $C$ of the channel, here they indicate the party to whom the underlying state belongs). We consider the shared state above to construct the first layer of our code. Conditioned on each and everyone of the $2^{R_{0}}$ states above, the parties are assumed to share $2^{R_{s}+R_{0}}$ copies of the state $\rho^{V^{A}V^{B}V^{C}}=\sum_{v}p(v)|vvv\rangle\langle vvv|^{V^{A}V^{B}V^{C}}$, as given below for the $i$-th $\rho^{U^{A}U^{B}U^{C}}$ state:
\begin{align*}
\sum_{u}p(u)|uuu\rangle\langle uuu|^{U^{A}_{i}U^{B}_{i}U^{C}_{i}}\otimes(\rho^{V^{A}V^{B}V^{C}}_{u})^{\otimes 2^{R_{s}+R_{1}}},
\end{align*}
%\begin{align*}
%=\rho^{U^{A}_{i}U^{B}_{i}U^{C}_{i}}\otimes(\sum_{v}p(v|u)|v\rangle\langle v|^{V^{A}_{1}}\otimes|v\rangle\langle v|^{V^{B}_{1}}\otimes|v\rangle\langle v|^{V^{C}_{1}})\otimes...\otimes(\sum_{v}p(v|u)|v\rangle\langle v|^{V^{A}_{2^{R_{s}+R_{1}}}}\otimes|v\rangle\langle v|^{V^{B}_{2^{R_{s}+R_{1}}}}\otimes|v\rangle\langle v|^{V^{C}_{2^{R_{s}+R_{1}}}}),
%\end{align*}
%We further define the $\rho^{V^{A}V^{B}V^{C}|U}\equiv\sum_{u}p(u)\rho^{V^{A}V^{B}V^{C}|U=u}$.
where Alice, Bob and Charlie are in possession of $V^{A},V^{B}$ and $V^{C}$ systems, respectively. The set $[1:2^{R_{s}+R_{1}}]$ is partitioned into $2^{R_{s}}$ equal size bins (and therefore inside each bin there are $2^{R_{1}}$ states). This constituted the second layer of the code. Finally for each and everyone of the states $\rho^{V^{A}V^{B}V^{C}}$, the parties will share $2^{R_{d}}$ copies of the state $\rho^{X^{A}X^{B}X^{C}}\coloneqq\sum_{x}p(x)|xxx\rangle\langle xxx|^{X^{A}X^{B}X^{C}}$, as mentioned below for the $i$-th state:
\begin{align*}
\sum_{u,v}p(u,v)|vvv\rangle\langle vvv|^{V^{A}_{i}V^{B}_{i}V^{C}_{i}}\otimes(\rho^{X^{A}X^{B}X^{C}}_{v})^{\otimes 2^{R_{d}}},
\end{align*}
where $X^{A},X^{B}$ and $X^{C}$ systems are owned by Alice, Bob and Charlie, respectively. These states build the third layer of the code. All states above are assumed to be available to all parties before communication begins. In the following, to avoid inefficient notation we may drop the superscripts if it does not lead to ambiguity; For instance when we analyze Bob's error probability, it is obvious that we are dealing with Bob's systems or in the secrecy analysis those of Charlie are dealt with.
\vspace*{5px}

\textbf{Encoding:} To send a message triple $(m_{0},m_{1},m_{s})$, the encoder first chooses a dummy message $m_{d}\in[1:2^{R_{d}}]$. In the first layer, the encoder finds the $m_{0}$-th state, i.e., $\rho^{U^{A}_{m_{0}}}$, then it looks for the $m_{s}$-th bin inside which, it selects the state associated to the private message $m_{1}$; Finally, the encoder picks the $m_{d}$-th state $\rho^{X^{A}_{m_{d}}}$ among those tied to the state found in the preceding step. The encoder sends the selected classical system through a modulator resulting in a quantum codeword $\rho^{A}_{x}$ which will be then transmitted over the channel\footnote{Note that we have included the modulator in the definition of the code meaning that it needs to be optimized over to get our capacity results.}. 
\vspace*{5px}

\textbf{Decoding:} Bob performs a two-phase decoding strategy such that he finds the common message in the first phase and then confidential and private messages in the subsequent phase. The transmission of the $m_{0}$-th common message induces the following state on Bob's side:
\begin{align}
\label{commoninduce}
\rho^{U_{1}}\otimes...\otimes\rho^{U_{m_{0}}B}\otimes...\otimes\rho^{U_{2^{R_{0}}}}.
\end{align}
Apparently Bob has to be able to locate the spot in which the received system $B$ is tied to his $U$ system. In other words, he should be to able to distinguish between states induced for different values of the common message. Bob employs a position-based decoding to solve the raised $2^{R_{0}}$-ary hypothesis testing problem. On the other hand, for the common message $m_{0}$, the selection of the twin$(m_{s},m_{1})$ will induce the following state on Bob's side:
\begin{align}
\rho^{V_{(m_{0},1,1)}}\otimes...\otimes\rho^{V_{(m_{0},m_{s},m_{1})}B}\otimes...\otimes\rho^{V_{(m_{0},2^{R_{s}},2^{R_{1}})}}.
\end{align}
Bob runs his second position-based POVM to solve the $2^{R_{s}+R_{1}}$-ary hypothesis testing problem.
%\footnote{The reader may wonder how this does not sound to be going along the same lines as its classical counterpart; The thing is that our encoding scheme uses shared randomness, therefore at this stage for a given common message $m_{0}$, we cannot have a fixed realization $u$. This, however, holds after we derandomize the code.}. 
Charlie also runs his position-based decoding POVM to find out the transmitted common message. The state induced at his side comes about by replacing $B$ with $C$ in (\ref{commoninduce}).  
\vspace*{5px}

\textbf{Analysis of the probability of error:} We first analyze the error probability of the common message by studying Bob's first decoder and the error analysis of the Charlie can be carried out along the same lines. It is worth pointing out that although the messages encoded in the second layer might include dummy randomness, Bob will decode all of them and he can throw away the dummy messages after decoding. The dummy messages in the third layer will not be decoded.

Reconsider the state in (\ref{commoninduce}). To find out the transmitted common message, Bob has to be equipped with some discriminator such that he can distinguish between $2^{R_{0}}$ different states. As hinted before, this puts forward a $2^{R_{0}}$-ary hypothesis testing problem. Let $\{T^{UB},I-T^{UB}\}$ be the elements of a POVM that is chosen for discriminating between two states $\rho^{UB}$ and $\rho^{U}\otimes\rho^{B}$. Further, we assume that the test operator $T^{UB}$ decides correctly in favor of $\rho^{UB}$ with probability at least $1-(\varepsilon'-\delta_{1})$\footnote{For the sake of intelligibility, we choose to specify the error probability of the test operator to be $\varepsilon'-\delta_{1}$ to ensure that the error probability of the code will be larger than this and at most $\varepsilon'$.}. Bob will use the following square-root measurement to detect the common message:
\begin{align*}
\Omega_{m_{0}}\coloneqq\left(\sum_{m'_{0}=1}^{2^{R_{0}}}\Pi_{m'_{0}}\right)^{-\frac{1}{2}}\Pi_{m_{0}}\left(\sum_{m'_{0}=1}^{2^{R_{0}}}\Pi_{m'_{0}}\right)^{-\frac{1}{2}},
\end{align*} 
where $\Pi_{m_{0}}\coloneqq\mathbbm{1}^{U_{1}}\otimes...\otimes T^{U_{m_{0}}B}\otimes...\otimes\mathbbm{1}^{U_{2^{R_{0}}}}$ and $T^{U_{m_{0}}B}$ is the test operator. It can be easily checked that the set $\{\Omega_{m_{0}}\}_{m_{0}}$ constitutes a valid POVM, i.e., $\sum_{m_{0}}\Omega_{m_{0}}=\mathbbm{1}$. Besides, direct calculation shows that $\Tr\{\Pi_{m_{0}}(\rho^{U_{1}}\otimes...\otimes\rho^{U_{m_{0}}B}\otimes...\otimes\rho^{U_{2^{R_{0}}}})\}=\Tr\{\Pi_{m_{0}}\rho^{U_{m_{0}}B}\}$ and for any $m'_{0}\neq m_{0}$, $\Tr\{\Pi_{m_{0}}(\rho^{U_{1}}\otimes...\otimes\rho^{U_{m'_{0}}B}\otimes...\otimes\rho^{U_{2^{R_{0}}}})\}=\Tr\{\Pi_{m_{0}}(\rho^{U_{m_{0}}}\otimes\rho^{B})\}$.

Observe that the symmetric structure of the codebook generation and decoding triggers an average error probability that is equal to the individual error probabilities. Therefore, we assume $m_{0}=1$ was transmitted. Hence,
\begin{align*}
\text{Pr}(\hat{M}_{0}\neq1|M_{0}=1)&=\Tr\{(\mathbbm{1}-\Omega_{1})(\rho^{U_{1}B}\otimes...\otimes\rho^{U_{2^{R_{0}}}})\}\\
&\leq(1+c)\Tr\{(\mathbbm{1}-\Pi_{1})(\rho^{U_{1}B}\otimes...\otimes\rho^{U_{2^{R_{0}}}})\}\\
&\hspace*{1.5cm}+(2+c+c^{-1})\sum_{m_{0}\neq1}\Tr\{\Pi_{m_{0}}(\rho^{U_{1}B}\otimes...\otimes\rho^{U_{2^{R_{0}}}})\}\\
&\leq(1+c)(\varepsilon'-\delta_{1})+(2+c+c^{-1})2^{R_{0}-I^{\varepsilon'-\delta_{1}}_{H}(U;B)_{\rho}},
\end{align*}
where the first inequality follows from Lemma \ref{HN} and in the second inequality, the first term is based on the assumption and the second term follows from the definition of the hypothesis testing mutual information (see Definition \ref{htmi}). The last expression is set equal to $\varepsilon'$ and the optimal value of $c$ is derived as $c=\frac{\delta_{1}}{2\varepsilon'-\delta_{1}}$. Then, we will have
\begin{align*}
R_{0}=I^{\varepsilon'-\delta_{1}}_{H}(U;B)_{\rho}-\log_{2}(\frac{4\varepsilon'}{\delta_{1}^{2}}).
\end{align*}
In the same manner, it can be shown that the achievable rate of the common message to Charlie equals $R_{0}=I^{\varepsilon'-\delta_{2}}_{H}(U;C)_{\rho^{UC}}-\log_{2}(\frac{4\epsilon'}{\delta_{2}^{2}})$.

In an analogous way, the reliability analysis of the confidential and the private messages goes as follows. Consider a binary POVM with elements $\{Q^{UVB},\mathbbm{1}-Q^{UVB}\}$. The POVM is to discriminate the states $\rho^{UVB}=\sum_{u}p(u)|u\rangle\langle u|^{U}\otimes\rho^{VB}_{u}$ and $\rho^{V-U-B}\coloneqq\sum_{u}p(u)|u\rangle\langle u|^{U}\otimes\rho^{V}_{u}\otimes\rho^{B}_{u}$ such that the value of $Q^{UVB}$ estimates the state to be $\rho^{UVB}$. Assume the probability of failure to make a correct decision on $\rho^{UVC}$ is at most $\varepsilon'-\delta_{3}$, i.e., $\Tr\{(\mathbbm{1}-Q)\rho^{UVB}\}\leq\varepsilon'-\delta_{3}$. Bob will take the following square-root measurement POVM :
\begin{align*}
\Theta_{m_{s},m_{1}}\coloneqq\left(\sum_{m'_{s}=1}^{2^{R_{s}}}\sum_{m'_{1}=1}^{2^{R_{1}}}\Gamma_{m'_{s},m'_{1}}\right)^{-\frac{1}{2}}\Gamma_{m_{s},m_{1}}\left(\sum_{m'_{s}=1}^{2^{R_{s}}}\sum_{m'_{1}=1}^{2^{R_{0}}}\Gamma_{m'_{s},m'_{1}}\right)^{-\frac{1}{2}},
\end{align*} 
where $\Gamma_{m_{s},m_{1}}\coloneqq\mathbbm{1}^{V_{1,1}}\otimes...\otimes Q^{UV_{m_{s},m_{1}}B}\otimes...\otimes \mathbbm{1}^{V_{2^{R_{s}},2^{R_{1}}}}$ and $Q^{UV_{m_{s},m_{1}}B}$ is the binary test operator. Observe that $\sum_{m_{s},m_{1}}\Theta_{m_{s},m_{1}}=\mathbbm{1}$. It is easy to show that for all $m_{s},m_{1}$, we have $\Tr\{\Gamma_{m_{s},m_{1}}(\sum_{u}p(u)|u\rangle\langle u|^{U}\otimes\rho^{V_{1,1}}_{u}\otimes...\otimes\rho^{V_{m_{s},m_{1}}B}_{u}\otimes...\otimes\rho^{V^{B}_{2^{R_{s}},2^{R_{1}}}}_{u})\}=\Tr\{Q\rho^{UVB}\}$. On the other hand, for any $m'_{s}\neq m_{s}$ or $m'_{1}\neq m_{1}$, $\Tr\{\Gamma_{m_{s},m_{1}}(\sum_{u}p(u)|u\rangle\langle u|^{U}\otimes\rho^{V_{1,1}}_{u}\otimes...\otimes\rho^{V_{m'_{s},m'_{1}}B}_{u}\otimes...\otimes\rho^{V_{2^{R_{s}},2^{R_{1}}}}_{u})\}=\Tr\{Q\rho^{V-U-B}\}$. By the symmetry of the random codebook construction, the average error probability is the same as the error probability of any twin $(m_{s},m_{1})$, hence it suffices to find the error probability if $(m_{s}=1,m_{1}=1)$ was sent. The analysis continues as follows:
\begin{align*}
&\text{Pr}((\hat{M}_{s},\hat{M}_{1})\neq(1,1)|(M_{s},M_{1})=(1,1))\\
&=\Tr\{(\mathbbm{1}-\Theta_{1,1})(\sum_{u}p(u)|u\rangle\langle u|^{U}\otimes\rho^{V_{1,1}B}_{u}\otimes...\otimes\rho^{V_{2^{R_{s}},2^{R_{1}}}}_{u})\}\\
&\leq(1+c)\Tr\{(\mathbbm{1}-\Pi_{1})(\sum_{u}p(u)|u\rangle\langle u|^{U}\otimes\rho^{V_{1,1}B}_{u}\otimes...\otimes\rho^{V_{2^{R_{s}},2^{R_{1}}}}_{u})\}\\
&\hspace*{2cm}+(2+c+c^{-1})\sum_{m_{0}\neq1}\Tr\{\Pi_{m_{0}}(\sum_{u}p(u)|u\rangle\langle u|^{U}\otimes\rho^{V_{1,1}B}_{u}\otimes...\otimes\rho^{V_{2^{R_{s}},2^{R_{1}}}}_{u})\}\\
&\leq(1+c)(\varepsilon'-\delta_{3})+(2+c+c^{-1})2^{R_{0}+R_{s}-I^{\varepsilon'-\delta_{3}}_{H}(V;B|U)_{\rho^{UVB}}},
\end{align*} 
where the first inequality is due to Lemma \ref{HN} and in the second inequality, the first term comes from the assumption about the accuracy of the test operator $Q$ and the second term uses the definition of the hypothesis testing conditional mutual information, Definition \ref{htcma}. We choose the error probability be less that or equal to $\varepsilon'$, so the optimal value of the constant is set to $c=\frac{\delta_{3}}{2\varepsilon'-\delta_{3}}$ and eventually we will get the following sum rate:
\begin{align*}
R_{s}+R_{1}=I_{H}^{\varepsilon'-\delta_{3}}(V;B|U)_{\rho}-\log_{2}(\frac{4\varepsilon'}{\delta_{3}^{2}}).
\end{align*}

\vspace*{5px}

\textbf{Analysis of the secrecy:} Our tool to study secrecy is the conditional convex-split lemma. The dummy message and perhaps private message which take care of confidentiality are encoded in the second and third layers as superposition of shared states. The quantum channel resolvability via superposition coding was studied in \cite{Anurag-Hayashi-Warsi.2018}. Given the setup of our problem, here we should try to prove the resolvability problem using convex-split lemma. We gave the analysis for Charlie's successful detection of the common message, hence in the secrecy analysis we assume Charlie knows the common message and so the correct copy of the $\rho^{U}$ used in the first layer. The idea for secrecy is that Charlie's systems have to remain close to some constant state, no matter which confidential message was transmitted.

For a given confidential message, the choice of the private message will induce an average state on Charlie's $V$ systems in the second layer where the dummy message induces an average state on his $X$ systems in the third layer. Since the states in the second layer are superposed to those in the third layer, both the private message and the dummy message will help to induce a state at Charlie's side that should be close enough to a target state. For a particular choice of the dummy message $m_{d}\in[1:2^{R_{d}}]$, the induced state at Charlie's side will be as follows:
\begin{align*}
\Psi^{C}_{v}\coloneqq\frac{1}{2^{R_{d}}}\sum_{i=1}^{2^{R_{d}}}\rho^{X^{C}_{1}}_{v}\otimes...\otimes\rho^{X^{C}_{i}C}_{v}\otimes...\otimes\rho^{X^{C}_{2^{R_{d}}}}_{v}.
\end{align*}
On the other hand, as mentioned, the private message also has its own share in the induced average state at Charlie where it influences the states in the second layer; More precisely, for a pick of the confidential message, we have
\begin{align*}
\sum_{u}p(u)|u\rangle\langle u|^{U}\otimes\bigg(\sum_{v}p(v|u)|v&\rangle\langle v|^{V_{1,1}}\otimes(\rho^{X}_{v})^{\otimes2^{R_{d}}}\otimes...\otimes\sum_{v}p(v|u)|v\rangle\langle v|^{V_{m_{s}-1,2^{R_{1}}}}\otimes(\rho^{X}_{v})^{\otimes2^{R_{d}}}\otimes
\frac{1}{2^{R_{1}}}\sum^{2^{R_{1}}}_{j=1}\Upsilon^{C,j}_{u}\\
&\otimes\big(\sum_{v}p(v|u)|v\rangle\langle v|^{V_{m_{s}+1,1}}\otimes(\rho^{X}_{v})^{\otimes2^{R_{d}}}\big)\otimes...\otimes\big(\sum_{v}p(v|u)|v\rangle\langle v|^{V_{2^{R_{s}},2^{R_{1}}}}\otimes(\rho^{X}_{v})^{\otimes2^{R_{d}}}\big)\bigg).
\end{align*} 
where
\begin{align*}
\Upsilon^{C,j}_{u}\coloneqq\sum_{v}p(v|u)|v\rangle\langle v|^{V_{m_{s},1}}\otimes(\rho^{X}_{v})^{\otimes 2^{R_{d}}}\otimes...\otimes\sum_{v}p(v|u)|v\rangle\langle v|^{V_{m_{s},j}}\otimes\Psi^{C}_{v}
\otimes...\otimes\sum_{v}p(v|u)|v\rangle\langle v|^{V_{m_{s},2^{R_{1}}}}\otimes(\rho^{X}_{v})^{\otimes 2^{R_{d}}}
\end{align*}
Charlie not being able to crack the confidential message amounts to his states being sufficiently close to the following state:
\begin{align}
\nonumber
\sum_{u}p(u)|u\rangle\langle u|^{U}\otimes&\Big(\big(\sum_{v}p(v|u)|v\rangle\langle v|^{V_{1,1}}\otimes(\rho^{X}_{v})^{\otimes2^{R_{d}}}\big)
\otimes...\otimes\big(\sum_{v}p(v|u)|v\rangle\langle v|^{V_{m_{s}-1,2^{R_{1}}}}\otimes(\rho^{X}_{v})^{\otimes2^{R_{d}}}\big)\\
\label{bracketterm}
&\otimes\bigg[\big(\sum_{v}p(v|u)|v\rangle\langle v|^{V_{m_{s}}}\otimes(\rho^{X^{C}}_{v})^{\otimes 2^{R_{d}}}\big)^{\otimes 2^{R_{1}}}\otimes\rho^{C}_{u}\bigg]
\otimes...\otimes\big(\sum_{v}p(v|u)|v\rangle\langle v|^{V_{R_{s},R_{1}}}\otimes(\rho^{X}_{v})^{2^{R_{d}}}\big)\Big).
\end{align}
where $\rho^{C}_{u}=\sum_{v,x}p(v,x|u)\rho^{C}_{x}$ is considered the constant state independent of the chosen confidential message. Concerning the trace distance between the aforementioned states, since the trace distance is invariant with respect to tensor product states, we can remove the same terms from both states. Eventually the following is what we want to be small enough:
\begin{align}
\label{secr1}
\frac{1}{2}\big\|\sum_{u}p(u)|u\rangle\langle u|^{U}\otimes\frac{1}{2^{R_{1}}}\sum^{2^{R_{1}}}_{j=1}\Upsilon^{C,j}_{u}-\sum_{u}p(u)|u\rangle\langle u|^{U}\otimes\big(\sum_{v}p(v|u)|v\rangle\langle v|^{V}\otimes(\rho^{X}_{v})^{\otimes 2^{R_{d}}}\big)^{\otimes 2^{R_{1}}}\otimes\rho^{C}_{u}\big\|_{1},
\end{align}
where the expression being subtracted refers to the state associated to the chosen confidential message given inside the brackets in (\ref{bracketterm}).
We proceed to show the above inequality by envisioning an intermediate state which is, intuitively, closer to either of the states involved in (\ref{secr1}) than the two states themselves. We define such an intermediate state as $\sum_{u}p(u)|u\rangle\langle u|^{U}\otimes\Xi^{C}_{u}$ where
\begin{align*}
\Xi^{C}_{u}\coloneqq\frac{1}{2^{R_{1}}}&\sum_{j=1}^{2^{R_{1}}}\bigg(\big[\sum_{v}p(v|u)|v\rangle\langle v|^{V_{m_{s},1}}\otimes(\rho^{X}_{v})^{\otimes 2^{R_{d}}}\big]\otimes...\\
&\hspace*{1cm}\otimes\big[\sum_{v}p(v|u)|v\rangle\langle v|^{V_{m_{s},j}}\otimes(\rho^{X_{1}}_{v}\otimes...\otimes\rho^{X_{R_{d}}}_{v}\otimes\rho^{C}_{v})\big]\otimes...\otimes\big[\sum_{v}p(v|u)|v\rangle\langle v|^{V_{m_{s},2^{R_{1}}}}\otimes(\rho^{X}_{v})^{\otimes 2^{R_{d}}}\big]\bigg).
\end{align*}

 Next, we have to bring in the intermediate state. We do so by the triangle inequality as follows 
\begin{align*}
&\frac{1}{2}\big\|\sum_{u}p(u)|u\rangle\langle u|^{U}\otimes\frac{1}{2^{R_{1}}}\sum^{2^{R_{1}}}_{j=1}\Upsilon^{C,j}_{u}-\sum_{u}p(u)|u\rangle\langle u|^{U}\otimes\big(\sum_{v}p(v|u)|v\rangle\langle v|^{V}\otimes(\rho^{X}_{v})^{\otimes 2^{R_{d}}}\big)^{\otimes 2^{R_{1}}}\otimes\rho^{C}_{u}\big\|_{1}\\
&\hspace*{2cm}\leq\frac{1}{2}\big\|\sum_{u}p(u)|u\rangle\langle u|^{U}\otimes\frac{1}{2^{R_{1}}}\sum^{2^{R_{1}}}_{j=1}\Upsilon^{C,j}_{u}-\sum_{u}p(u)|u\rangle\langle u|^{U}\otimes\Xi^{C}_{u}\|_{1}\\
&\hspace*{6cm}+\frac{1}{2}\big\|\sum_{u}p(u)|u\rangle\langle u|^{U}\otimes\Xi^{C}_{u}-\sum_{u}p(u)|u\rangle\langle u|^{U}\otimes\big(\rho^{V^{C}}_{u}\otimes(\rho^{X^{C}}_{v})^{\otimes 2^{R_{d}}}\big)^{\otimes 2^{R_{1}}}\otimes\rho^{C}_{u}\big\|_{1}.
\end{align*}
We now try to upper bound each term appeared on the right-hand side. For the first term, simply by expanding the summation and subtracting equal terms from both side, it can be seen that:
\begin{align*}
\frac{1}{2}\big\|\sum_{u}p(u)&|u\rangle\langle u|^{U}\otimes(\frac{1}{2^{R_{1}}}\sum^{2^{R_{1}}}_{j=1}\Upsilon^{C,j}_{u}-\Xi^{C}_{u})\big\|_{1}\\
&=\frac{1}{2}\sum_{u}p(u)\Big\|\sum_{v}p(v|u)|v\rangle\langle v|^{V}\otimes\big(\rho^{X_{1}}_{v}\otimes...\otimes\rho^{X_{2^{R_{d}}}}_{v}\otimes\rho^{C}_{v}-\sum_{i=1}^{2^{R_{d}}}\rho^{X}_{v}\otimes...\otimes\rho^{X_{i}C}_{v}\otimes...\otimes\rho^{X_{2^{R_{d}}}}_{v}\big)\Big\|_{1}.
\end{align*}
Then immediately by noting the Markov chain, the conditional convex-split lemma  asserts that if $R_{d}=\widetilde{I}^{\varepsilon''}_{\ma}(X;C|V)_{\rho}+2\log_{2}(\frac{1}{\eta})$, then
\begin{align*}
\frac{1}{2}\Big\|\sum_{u}p(u)&|u\rangle\langle u|^{U}\otimes(\frac{1}{2^{R_{1}}}\sum^{2^{R_{1}}}_{j=1}\Upsilon^{C,j}_{u}-\Xi^{C}_{u})\Big\|_{1}\leq2\epsilon''+\eta,
\end{align*}
and from the relation between the purified distance and the trace distance, we have 
\begin{align*}
P(\sum_{u}p(u)|u\rangle\langle u|^{U}\otimes\frac{1}{2^{R_{1}}}\sum^{2^{R_{1}}}_{j=1}\Upsilon^{C,j}_{u},\sum_{u}p(u)|u\rangle\langle u|^{U}\otimes\Xi^{C}_{u})\leq2\varepsilon''+\eta.
\end{align*}
For the second term, from the invariance of the trace distance with respect to tensor product states, we can trace out $X$ systems from both expressions leading to the following:
\begin{align*}
\frac{1}{2}\bigg\|\sum_{u}p(u)|u\rangle\langle u|^{U}\otimes\big(\frac{1}{2^{R_{1}}}\sum_{j=1}^{R_{1}}(\rho^{V^{C}_{1}}_{u}\otimes...\otimes\rho^{V^{C}_{i}C}_{u}\otimes...\otimes\rho^{V^{C}_{R_{1}}}_{u})-\rho^{V^{C}_{1}}_{u}\otimes...\otimes\rho^{V^{C}_{2^{R_{1}}}}_{u}\otimes\rho^{C}_{u}\big)\bigg\|_{1},
\end{align*}
then the conditional convex-split lemma guarantees the above to be less that or equal to $(2\varepsilon''+\eta)$ if we choose $R_{1}=\widetilde{I}_{\ma}^{\varepsilon''}(V;C|U)_{\rho}+2\log_{2}(\frac{1}{\eta})$, which in turn, implies that the purified distance between the target states is also less that or equal to $(2\varepsilon''+\eta)$.

\vspace*{5px}

\textbf{Derandomizarion:} Our protocol so far has relied upon shared randomness among parties. In order to show that the results also hold without assistance of shared randomness, we need to derandomize the code. The derandomization is a standard procedure which can be done by expanding the states and corresponding POVMS and using a property of the trace distance given by the equality in (\ref{derandomizationaa}) (see  \cite{Farzin.arXiv.2018}, \cite{Mark2017}, \cite{Mark-app}). The only point that might be needed to be made here is the structure of the test operators in Bob's decoders (as well as that of Charlie). Note than the test operators were described generally as $T^{UB}$ and $Q^{UVB}$ without specifying the nature of the subsystems, i.e., whether each of $U,V$ or $B$ systems are classical or quantum. For our purposes, it is sufficient to consider the test operators as $T^{UB}\coloneqq\sum_{u}|u\rangle\langle u|^{U}\otimes \overline{T}_{u}^{B}$ where $\overline{T}_{u}^{B}\coloneqq\langle u|T^{UB}|u\rangle$. Likewise, we only need to have $Q^{UVB}\coloneqq\sum_{u,v}|u\rangle\langle u|^{U}\otimes|v\rangle\langle v|^{V}\otimes\overline{Q}^{B}_{u,v}$ where $\overline{Q}^{B}_{u,v}\coloneqq\langle u,v|Q^{UVB}|v,u\rangle$.
\vspace*{5px}

\textbf{Expurgation:} So far we have come to know that there exists at least one code that satisfies the reliability criterion in (\ref{error}) and at least one codebook that satisfies the secrecy requirement (\ref{secrecy}). We should use Markov inequality to find a good code that satisfies both the reliablity (\ref{error}) and secrecy (\ref{secrecy}) simultaneouslly. Moreover, from the gentle measurement lemma \cite{Gentlem-1999} we know that the disturbed state fed into the second decoder of Bob is impaired by at most $2\sqrt{\epsilon'}$. we have the average error probability over all codes $P_{\text{error}}^{1}\leq3\epsilon'+2\sqrt{\epsilon'}$ and the secrecy over all code $P_{\text{secrecy}}^{1}\leq4\epsilon'+2\eta$. From Markov inequality we know that $\text{Pr}(P_{\text{error}}^{1}\geq\sqrt[4]{\epsilon'})\leq3(\epsilon')^{3/4}+2\sqrt[4]{\epsilon'}$ and $\text{Pr}(P_{\text{secrecy}}^{1}\geq\sqrt[4]{\epsilon''})\leq4(\epsilon'')^{3/4}+2(\epsilon'')^{7/4}$. Then there is a good code that with high probability neither statement is true:
{\small{
\begin{align*}
&\text{Pr}(P_{\text{error}}^{1}\leq\sqrt[4]{\epsilon'},P_{\text{secrecy}}^{1}\leq\sqrt[4]{\epsilon''})\\
&\hspace*{1.5cm}\geq 1-(3(\epsilon')^{3/4}+2\sqrt[4]{\epsilon'})-(4(\epsilon'')^{3/4}+2(\epsilon'')^{7/4}).
\end{align*}
}}
Let $\epsilon\coloneqq \text{max}\{\sqrt[4]{\epsilon'},\sqrt[4]{\epsilon''}\}$. This parameter works for both requirements and the results is concluded.
\section{Converse}
We go over the bounds one at a time. Consider the reliability of the common message ($\ref{commonc}$). From the definition of the reliability (\ref{error}), the union bound suggests that $\text{Pr}\{\hat{M}_{0}\neq M_{0}\}\leq\varepsilon$ and $\text{Pr}\{\tilde{M}_{0}\neq M_{0}\}\leq\varepsilon$. Then converse bound (\ref{commonc}) has shown in \cite{Wang-Renner} by relating the communication problem to a problem in binary hypothesis testing. We briefly explain the approach here. Consider the task of distinguishing between two hypothesis $\rho^{\hat{M}_{0}M_{0}}=\frac{1}{2^{R_{0}}}\sum_{m_{0}}\ketbra{m_{0}}^{\hat{M}_{0}}\otimes\ketbra{m_{0}}^{M_{0}}$ and $\rho^{\hat{M}_{0}}\otimes\rho^{M_{0}}$ where the former is the null hypothesis and the latter null hypothesis. It can be easily verified that $\text{Pr}\{\hat{M}_{0}\neq M_{0}\}\leq\varepsilon$ implies that the type I error is less that or equal to $\varepsilon$ and the type II error equals $2^{-R_{0}}$. Then from the definition of the hypothesis-testing mutual information and the monotonicty of the hypothesis testing relative entropy with CPTP maps, we have $R_{0}\leq I_{\h}^{\varepsilon}(M_{0};B)_{\rho}$. Let $U\coloneqq M_{0}$, then the converse follows. The proof of $R_{0}\leq I_{\h}^{\varepsilon}(M_{0};C)_{\rho}$ follows the same argument.

From (\ref{error}), the union of events imply both $\Pr\{(\hat{M}_{0},\hat{M}_{1},\hat{M}_{s})\neq(M_{0},M_{1},M_{s})\}\leq\varepsilon$ and $\Pr\{\tilde{M}_{0}\neq M_{0}\}\leq\varepsilon$. Similar to \cite{Farzin.arXiv.2018}, we expand the former as follows:
\begin{align*}
\varepsilon&\geq\Pr\{(\hat{M}_{0},\hat{M}_{1},\hat{M}_{s})\neq(M_{0},M_{1},M_{s})\}\\
&=\sum_{m_{0},m_{1},m_{s}}p(m_{0})p(m_{1})p(m_{s})\Pr\{(\hat{M}_{0},\hat{M}_{1},\hat{M}_{s})\neq(m_{0},m_{1},m_{s})|m_{0},m_{1},m_{s}\}\\
&=\sum_{m_{0},m_{1},m_{s}}p(m_{0})p(m_{1})p(m_{s})\sum_{(m'_{0},m'_{1},m'_{s})\neq(m_{0},m_{1},m_{s})}p(m'_{0},m'_{1},m'_{s}|m_{0},m_{1},m_{s})\\
&\geq\sum_{m_{0},m_{1},m_{s}}p(m_{0})p(m_{1})p(m_{s})\sum_{\substack{m'_{0},\\(m'_{1},m'_{s})\neq(m_{1},m_{s})}}p(m'_{0},m'_{1},m'_{s}|m_{0},m_{1},m_{s})\\
&=\sum_{m_{0},m_{1},m_{s}}p(m_{0})p(m_{1})p(m_{s})\sum_{\substack{(m'_{1},m'_{s})\neq(m_{1},m_{s})}}p(m'_{1},m'_{s}|m_{0},m_{1},m_{s})\\
&=\sum_{m_{0}}p(m_{0})\Pr\{(\hat{M}_{1},\hat{M}_{s})\neq (M_{1},M_{s})|M_{0}=m_{0}\}.
\end{align*}  
Notice that the final expression indicates the probability of erroneous detection of $(M_{s},M_{1})$ while $M_{0}$ is known. We find an upper bound on the sum rate of $(M_{s},M_{1})$ by considering a binary hypothesis testing problem with null and alternative hypotheses given respectively as follows:
\begin{align*}
\rho^{M_{0}\hat{M_{s}}\hat{M_{1}}M_{s}M_{1}}&\coloneqq\frac{1}{2^{R_{0}}}\sum_{m_{0}}\ketbra{m_{0}}^{M_{0}}\otimes\rho^{\hat{M_{s}}\hat{M_{1}}M_{s}M_{1}}_{m_{0}},\\
\rho^{\hat{M_{s}}\hat{M_{1}}-M_{0}-M_{s}M_{1}}&\coloneqq\frac{1}{2^{R_{0}}}\sum_{m_{0}}\ketbra{m_{0}}^{M_{0}}\otimes\rho^{\hat{M_{s}}\hat{M_{1}}}_{m_{0}}\otimes\rho^{M_{s}M_{1}}_{m_{0}},
\end{align*}
where $\rho^{\hat{M_{s}}\hat{M_{1}}M_{s}M_{1}}_{m_{0}}=\frac{1}{2^{R_{s}+R_{1}}}\sum_{m_{s}m_{1}}\ketbra{m_{s}m_{1}}^{\hat{M_{s}}\hat{M_{1}}}\otimes\ketbra{m_{s}m_{1}}^{M_{s}M_{1}}$. It can be easily verified that type I error is equivalent to $\sum_{m_{0}}p(m_{0})\Pr\{(\hat{M}_{1},\hat{M}_{s})\neq (M_{1},M_{s})|M_{0}=m_{0}\}$ which was shown to be less that or equal to $\varepsilon$. On the other hand, the type II error can be written as follows:
\begin{align*}
\sum_{m_{0},m_{s},m_{1}}p_{M_{0}}(m_{0})p_{M_{s}M_{1}}(m_{s},m_{1})p_{\hat{M}_{s}\hat{M}_{1}}(m_{s},m_{1})=\frac{1}{2^{R_{s}+R_{1}}}\sum_{m_{0},m_{s},m_{1}}p_{M_{0}}(m_{0})p_{\hat{M}_{s}\hat{M}_{1}}(m_{s},m_{1})=\frac{1}{2^{R_{s}+R_{1}}}.
\end{align*}   
Then we have the following:
\begin{align*}
R_{s}+R_{1}\leq I_{\h}^{\varepsilon}(M_{s},M_{1};\hat{M}_{s},\hat{M}_{1}|M_{0})_{\rho}\leq  I_{\h}^{\varepsilon}(M_{s},M_{1};B|M_{0})_{\rho},
\end{align*}
where the first inequality stems from the definition of the conditional hypothesis testing mutual information and the second inequality is from monotonicity under CPTP maps. Identifying the random variables $V\coloneqq(M_{s},M_{1})$ and $U\coloneqq M_{0}$ concludes the intended bound. So far we have dealt with the reliability condition and have derived (\ref{commonc}) and (\ref{allc}). 

Next we turn our attention to the secrecy criterion. The secrecy condition (\ref{privacy}) requires that the state of the Charlie and the confidential message become close to a product state for every transmitted common message. In converse proof, we consider a less strict criterion such that we demand the aforementioned states to be close on average over the common messages. i.e., 
\begin{align*}
\frac{1}{2}\big\|\rho^{CM_{0}M_{s}}-\frac{1}{2^{R_{0}}}\sum_{m_{0}}\ketbra{m_{0}}^{M_{0}}\otimes\rho^{M_{s}}_{m_{0}}\otimes\sigma^{C}_{m_{0}}\big\|=\frac{1}{2^{R_{0}+R_{s}}}\sum_{m_{0},m_{s}}\frac{1}{2}\|\rho^{C}_{m_{0},m_{s}}-\sigma^{C}_{m_{0}}\|_{1}\leq\varepsilon.
\end{align*} 
From the relation between the purified distance and the trace distance, the purified distance between the above-mentioned states is less that or equal to $\sqrt{2\varepsilon}$. Then from the definition of the smooth conditional relative entropy, it is easily checked that the following holds:
\begin{align*}
D_{\ma}^{\sqrt{2\varepsilon}}(M_{s};C|M_{0})_{\rho}\coloneqq D_{\ma}^{\sqrt{2\varepsilon}}\big(\rho^{M_{0}M_{s}C}\big\|\frac{1}{2^{R_{0}}}\sum_{m_{0}}\ketbra{m_{0}}^{M_{0}}\otimes\rho^{M_{s}}_{m_{0}}\otimes\sigma^{C}_{m_{0}}\big)_{\rho}=0.
\end{align*} 
Therefore, in the quantity $D_{\ma}^{\sqrt{2\varepsilon}}(M_{s};C|M_{0})_{\rho}=0$, we define $U\coloneqq M_{0}$ and $V\coloneqq M_{s}$ to get $D_{\ma}^{\sqrt{2\varepsilon}}(V;C|U)_{\rho}=0$. Similarly, we let $V\coloneqq M_{0}$ and  $X\coloneqq M_{s}$ to get $D_{\ma}^{\sqrt{2\varepsilon}}(X;C|V)_{\rho}=0$. Finally the bound on the rate of the confidential message (\ref{confic}), can be seen from the bound derived on $R_{s}+R_{1}$ and the preceding discussion.
\section{Asymptotic Analysis}
So far we have studied the scenario in which a quantum channel is available only once and the transmission was subject to some non-zero error and secrecy parameters. In the asymptotic regime, however, a memoryless channel is considered to be available for an unlimited number of uses; If we denote the uses of the channel by $n$, the one-shot scenario corresponds to $n=1$ where in the asymptotic regime $n\rightarrow \infty$. Moreover, in the asymptotic regime as long as the achievability bounds and weak converses are concerned, the error and secrecy parameters are assumed to be vanishing in the limit of many channel uses, i.e., $\varepsilon\rightarrow 0$ as $n\rightarrow \infty$. The following formally defines the rate region in the asymptotic regime from the one-shot rate region defined before:
\begin{align}
\label{asymdef}
\mathcal{R}_{\infty}(\mathcal{N})\coloneqq\lim_{\varepsilon\rightarrow 0}\lim_{n\rightarrow\infty}\frac{1}{n}\mathcal{R}^{\varepsilon}(\mathcal{N}^{\otimes n}),
\end{align}
where $\mathcal{N}^{\otimes n}$ indicates the $n$ independent uses of the channel $\mathcal{N}$.
In the following we first prove a theorem then we will recover several well-known results as corollaries.
\begin{theorem}
\label{asyt}
The asymptotic rate region $\mathcal{R}_{\infty}(\mathcal{N})$ of the broadcast channel $\mathcal{N}^{A\rightarrow BC}$ is given as follows:
\begin{align}
\label{regionasym}
\mathcal{R}_{\infty}(\mathcal{N})=\bigcup_{\ell=1}^{\infty}\frac{1}{\ell}\mathcal{R}_{\infty}^{(1)}(\mathcal{N}^{\otimes\ell}),
\end{align}
where $\mathcal{R}_{\infty}^{(1)}(\mathcal{N})\coloneqq\bigcup_{\rho^{UVXBC}}\mathcal{R}_{\infty}^{(2)}(\mathcal{N})$, in which $\mathcal{R}_{\infty}^{(2)}(\mathcal{N})$  is the set of quadruples $(R_{0},R_{1},R_{s},R_{d})$ satisfying the following conditions:
\begin{align}
\label{commonaa}
R_{0}&\leq\text{min}\big[I(U;B)_{\rho},I(U;C)_{\rho}\big], \\
\label{allaa}
R_{0}+R_{1}+R_{s}&\leq I(V;B|U)_{\rho}+\text{min}\big[I(U;B)_{\rho},I(U;C)_{\rho}\big], \\
\label{confiaa}
R_{s}&\leq I(V;B|U)_{\rho}-I(V;C|U)_{\rho}, \\
\label{convex2aa}
R_{1}+R_{d}&\geq I(V;C|U)_{\rho}+I(X;C|V)_{\rho},\\
\label{convex1aa}
R_{d}&\geq I(X;C|V)_{\rho},
\end{align}
where $\rho^{UVXBC}=\sum_{u,v,x}p(u,v)p(x|v)\ketbra{u}^{U}\otimes\ketbra{v}^{V}\otimes\ketbra{x}^{X}\otimes\mathcal{N}(\rho_{x}^{A})$ is the state arising from the channel.
\end{theorem}

%\begin{remark}
%Note that $\mathcal{R}_{\infty}^{1}(\mathcal{N}^{\otimes\ell})=\bigcup_{\rho^{\ell}}\mathcal{R}_{\infty}^{2}(\mathcal{N}^{\otimes\ell})$ and $\mathcal{R}_{\infty}^{2}(\mathcal{N}^{\otimes\ell})$ consists of the rate quadruples $(R_{0},R_{1},R_{s},R_{d})$ satisfying the following: 
%\begin{align*}
%R_{0}&\leq\text{min}\big[I(U^{\ell};B^{\otimes\ell})_{\rho},I(U^{\ell};C^{\otimes\ell})\big], \\
%R_{0}+R_{1}+R_{s}&\leq I(V^{\ell};B^{\otimes\ell}|U^{\ell})+\text{min}\big[I(U^{\ell};B^{\otimes\ell})_{\rho^{UB}},I(U^{\ell};C^{\otimes\ell})_{\rho^{UC}}\big], \\
%R_{s}&\leq I(V^{\ell};B^{\otimes\ell}|U^{\ell})-I(V^{\ell};C^{\otimes\ell}|U^{\ell}), \\
%R_{1}+R_{d}&\geq I(V^{\ell};C^{\otimes\ell}|U^{\ell})+I(X^{\ell};C^{\otimes\ell}|V^{\ell}),\\
%R_{d}&\geq I(X^{\ell};C^{\otimes\ell}|V^{\ell}).
%\end{align*}
%for a state $\rho^{\ell}$ arising from $\ell$ uses of the channel such that $U^{\ell},V^{\ell}$ and $X^{\ell}$ refer to classical random variables on $\mathcal{U}^{\ell},\mathcal{V}^{\ell}$ and $\mathcal{X}^{\ell}$, respectively and $B^{\otimes\ell}$ and $C^{\otimes\ell}$ refer to the $n$-fold tensor product of the Hilbert spaces $\mathcal{H}^{B}$ and $\mathcal{H}^{C}$, respectively.
%\end{remark}
\begin{IEEEproof}[Proof of Theorem \ref{asyt}]
We need to show the direct part and the converse. To establish the direct part, we appeal to our one-shot achievability region and seek to show that the right-hand side of equation (\ref{regionasym}) is contained inside the left-hand side, i.e., the following:
\begin{align*}
\bigcup_{\ell=1}^{\infty}\frac{1}{\ell}\mathcal{R}_{\infty}^{(1)}(\mathcal{N}^{\otimes\ell})\subseteq\mathcal{R}_{\infty}(\mathcal{N}).
\end{align*}
From our achievability result Theorem \ref{achievability}, if we use the channel $m$ times independently (memoryless channel), or equivalently if we consider one use of a \enquote{big channel} $\mathcal{N}^{\otimes m}$, we will have: 
\begin{align}
\label{direct}
\bigcup_{\rho^{m}}\mathcal{R}^{(in)}(\mathcal{N}^{\otimes m})\subseteq\mathcal{R}^{\varepsilon}(\mathcal{N}^{\otimes m}),
\end{align}
where $\mathcal{R}^{(in)}(\mathcal{N}^{\otimes m})$ is the convex closure over all states $\rho^{m}$ arising from $m$ uses of the channel, of the rate quadruples $(R_{0},R_{1},R_{s},R_{d})$ obeying the following:
\begin{align*}
R_{0}&\leq\text{min}\big[I^{\varepsilon'-\delta_{1}}_{\h}(U^{m};B^{\otimes m})_{\rho^{m}}-\log_{2}(\frac{4\varepsilon'}{\delta_{1}^{2}}),I_{\h}^{\varepsilon'-\delta_{2}}(U^{m};C^{\otimes m})_{\rho^{m}}-\log_{2}(\frac{4\varepsilon'}{\delta_{2}^{2}})\big], \\
R_{0}+R_{1}+R_{s}&\leq I_{\h}^{\varepsilon'-\delta_{3}}(V^{m};B^{\otimes m}|U^{m})_{\rho^{m}}-\log_{2}(\frac{4\varepsilon'}{\delta_{3}^{2}})\\ \nonumber
&\hspace*{2cm}+\text{min}\big[I^{\varepsilon'-\delta_{1}}_{\h}(U^{m};B^{\otimes m})_{\rho^{m}}-\log_{2}(\frac{4\varepsilon'}{\delta_{1}^{2}}),I_{\h}^{\varepsilon'-\delta_{2}}(U^{m};C^{\otimes m})_{\rho^{m}}-\log_{2}(\frac{4\varepsilon'}{\delta_{2}^{2}})\big], \\
R_{s}&\leq I_{\h}^{\varepsilon'-\delta_{3}}(V^{m};B^{\otimes m}|U^{m})_{\rho^{m}}-\widetilde{I}_{\ma}^{\varepsilon''}(V^{m};C^{\otimes m}|U^{m})_{\rho^{m}}-\log_{2}(\frac{4\varepsilon'}{\delta_{1}^{2}})-2\log_{2}(\frac{1}{\eta}), \\
R_{1}+R_{d}&\geq\widetilde{I}_{\ma}^{\varepsilon''}(V^{m};C^{\otimes m}|U^{m})_{\rho^{m}}+\widetilde{I}_{\ma}^{\varepsilon''}(X^{m};C^{\otimes m}|V^{m})_{\rho^{m}}+4\log_{2}(\frac{1}{\eta}),\\
R_{d}&\geq \widetilde{I}_{\ma}^{\varepsilon''}(X^{m};C^{\otimes m}|V^{m})_{\rho^{m}}+2\log_{2}(\frac{1}{\eta}),
\end{align*}
where $U^{m},V^{m}$ and $X^{m}$ refer to the random variables drawn from the joint distributions $p(u_{1},...,u_{m}), p(v_{1},...,v_{m})$ and $p(x_{1},...,x_{m})$, respectively and $B^{\otimes m}$ and $C^{\otimes m}$ refer to the $m$-fold tensor product of the Hilbert spaces $\mathcal{H}^{B}$ and $\mathcal{H}^{C}$, respectively. Since we are now after some achievability theorem, we can assume that each sequence of random variables is drawn from corresponding distributions in an i.i.d. fashion, i.e., for example $p(u_{1},...,u_{m})=\prod_{i=1}^{m}p(u_{i})$. Therefore, the state over which the above quantities are assessed, is $\rho^{\otimes m}=\rho\otimes...\otimes\rho$. 

The i.i.d. assumption enables us to simplify the entropic quantities in the asymptotic limit of many channel uses. To see this, we divide both sides of (\ref{direct}) by $m$ and let $m\rightarrow\infty$. This results in dividing the entropic quantities comprising $\mathcal{R}^{(in)}(\mathcal{N}^{\otimes m})$ by $m$ and evaluate limits as $m\rightarrow\infty$. All the constant terms will vanish as $m\rightarrow\infty$ and from the asymptotic i.i.d. behaviour of the quantities studied in Lemmas \ref{asym3} and Lemma \ref{asym4}, we get the region $\mathcal{R}^{1}_{\infty}(\mathcal{N})$. So far we have shown the following:
\begin{align*}
\mathcal{R}^{1}_{\infty}(\mathcal{N})\subseteq\lim_{\varepsilon\rightarrow 0}\lim_{m\rightarrow\infty}\frac{1}{m}\mathcal{R}^{\varepsilon}(\mathcal{N}^{\otimes m}),
\end{align*}
Finally we consider $m$ uses of the big channel $\mathcal{N}^{\otimes\ell}$ and let $n=m\ell$. Taking the limits as $n\rightarrow\infty$ concludes the direct part.

For the converse part, from Theorem \ref{converse1} onward, if the channel $\mathcal{N}$ gets used $n$ independent times, we will have 
\begin{align}
\label{containcon}
\mathcal{R}^{\varepsilon}(\mathcal{N}^{\otimes n})\subseteq\bigcup_{\ell=1}^{n}\bigcup_{\rho^{\ell}}\mathcal{R}^{(co)}(\mathcal{N}^{\otimes \ell}),
\end{align}
where $\mathcal{R}^{(co)}(\mathcal{N}^{\otimes \ell})$ consists of the rate quadruples $(R_{0},R_{1},R_{s},R_{d})$ obeying the following:
\begin{align*}
R_{0}&\leq\text{min}\big[I^{\varepsilon}_{\h}(U^{\ell};B^{\otimes \ell})_{\rho^{\ell}},I_{\h}^{\varepsilon}(U^{\ell};C^{\otimes \ell})_{\rho^{\ell}}\big], \\
R_{0}+R_{1}+R_{s}&\leq I_{\h}^{\varepsilon}(V^{\ell};B^{\otimes \ell}|U^{\ell})_{\rho^{\ell}}+\text{min}\big[I^{\varepsilon}_{\h}(U^{\ell};B^{\otimes \ell})_{\rho^{\ell}},I_{\h}^{\varepsilon}(U^{\ell};C^{\otimes \ell})_{\rho^{\ell}}\big], \\
R_{s}&\leq I_{\h}^{\varepsilon}(V^{\ell};B^{\otimes \ell}|U^{\ell})_{\rho^{\ell}}-D_{\ma}^{\sqrt{2\varepsilon}}(V^{\ell};C^{\otimes \ell}|U^{\ell})_{\rho^{\ell}}, \\
R_{1}+R_{d}&\geq D_{\ma}^{\sqrt{2\varepsilon}}(V^{\ell};C^{\otimes \ell}|U^{\ell})_{\rho^{\ell}}+D_{\ma}^{\sqrt{2\varepsilon}}(X^{\ell};C^{\otimes \ell}|V^{\ell})_{\rho^{\ell}},\\
R_{d}&\geq D_{\ma}^{\sqrt{2\varepsilon}}(X^{\ell};C^{\otimes \ell}|V^{\ell})_{\rho^{\ell}},
\end{align*}
where $(\rho^{UVXBC})^{\ell}$ is the state inducing by $\ell$ independent uses of the channel such that its classical systems, $U^{\ell},V^{\ell}$ and $X^{\ell}$ correspond to the random variables drawn from the joint distributions $p(u_{1},...,u_{\ell}), p(v_{1},...,v_{\ell})$ and $p(x_{1},...,x_{\ell})$, respectively and quantum systems $B^{\otimes \ell}$ and $C^{\otimes \ell}$ refer to the $\ell$-fold tensor product of the Hilbert spaces $\mathcal{H}^{B}$ and $\mathcal{H}^{C}$, respectively.
Each and everyone of the entropic quantities in the region above have been shown to be bounded by corresponding quantum relative entropies, see equations (\ref{relationH}), (\ref{relationHc}) and (\ref{relationMc}). By invoking the bounds, $\mathcal{R}^{(co)}(\mathcal{N}^{\otimes \ell})$ can be seen to be included in the following region: 
\begin{align*}
R_{0}&\leq\text{min}\big[I(U^{\ell};B^{\otimes \ell})_{\rho^{\ell}},I(U^{\ell};C^{\otimes \ell})_{\rho^{\ell}}\big], \\
R_{0}+R_{1}+R_{s}&\leq I(V^{\ell};B^{\otimes \ell}|U^{\ell})_{\rho^{\ell}}+\text{min}\big[I(U^{\ell};B^{\otimes \ell})_{\rho^{\ell}},I(U^{\ell};C^{\otimes \ell})_{\rho^{\ell}}\big], \\
R_{s}&\leq I(V^{n};B^{\otimes \ell}|U^{\ell})_{\rho^{\ell}}-I(V^{\ell};C^{\otimes \ell}|U^{\ell})_{\rho^{\ell}}, \\
R_{1}+R_{d}&\geq I(V^{\ell};C^{\otimes \ell}|U^{\ell})_{\rho^{\ell}}+I(X^{\ell};C^{\otimes \ell}|V^{\ell})_{\rho^{\ell}},\\
R_{d}&\geq I(X^{n};C^{\otimes \ell}|V^{\ell})_{\rho^{\ell}}.
\end{align*}
The proof will be completed by dividing both sides of (\ref{containcon}) by $n$ and letting $n\rightarrow\infty$ as well as $\varepsilon\rightarrow0$.
\end{IEEEproof}

\begin{corollary}[Theorem 1 in \cite{Devetak-Shor}]
%$\{p(u),\ket{\phi_{u}}\bra{\phi_{u}}^{RA}\}$ 
Consider the quantum channel $\mathcal{N}^{A\rightarrow B}$ with an isometric extension $V^{A\rightarrow BE}$ and let $\rho^{URA}=\sum_{u}p(u)\ketbra{u}\otimes\ket{\phi_{u}}\bra{\phi_{u}}^{RA}$ be a cq state in which $R$ is a reference system. The capacity region of simultaneous transmission of classical and quantum information for the channel is given by
\begin{align*}
\text{S}^{\infty}(\mathcal{N})=\bigcup_{\ell=1}^{\infty}\frac{1}{\ell}\text{S}_{1}^{\infty}(\mathcal{N}^{\otimes\ell}),
\end{align*} 
where $\text{S}_{1}^{\infty}(\mathcal{N})$ is the union, over all states of the form $\rho^{URB}=\sum_{u}p(u)\ketbra{u}\otimes\mathcal{N}^{A\rightarrow B}(\ket{\phi_{u}}\bra{\phi_{u}}^{RA})$ arising from the channel, of the rate pairs $(R_{c}^{\infty},R_{q}^{\infty})$ obeying:
\begin{align*}
R_{c}^{\infty}&\leq I(U;B)_{\rho},\\
R_{q}^{\infty}&\leq I(R\rangle BU)_{\rho},
\end{align*}
where $R_{c}^{\infty}$ and $R_{q}^{\infty}$ denote respectively the rates of the classical and quantum information and $I(R\rangle BU)_{\rho}\coloneqq-S(R|BU)_{\rho}$ is the \textit{coherent information}.
\begin{IEEEproof}
Following the discussion of Corollary \ref{simulone} and Theorem \ref{asyt}, we only need to argue that the coherent information of the ensemble $\{p(u),\ket{\phi_{u}}\bra{\phi_{u}}^{RBE}\}$ is equal to the rate of the confidential message in Theorem 3, i.e., the following:
\begin{align*}
I(R\rangle BU)_{\rho}=I(V;B|U)_{\rho}-I(V;E|U)_{\rho}.
\end{align*}
We apply the Schmidt decomposition to the pure states $\{\ket{\phi_{u}}^{RBE}\}_{u}$ with respect to the cut $R|BE$ and then measure the $R$ system in a suitable orthonormal basis. This measurement decoherifies the states such that the $R$ system can be shown by a classcial system, say $V$. Then the equality of the coherent information and the confidential message rate can be easily checked (see for example exercise 11.6.7 in \cite{Markbook}).
\end{IEEEproof}
\end{corollary}

\begin{corollary}[Theorem 3 of \cite{Wat-Ooh15}]
Let $\mathcal{N}_{C}^{X\rightarrow (Y,Z)}$ be a classical channel taking inputs to outputs according to some distribution $p(y,z|x)$. We define $\mathcal{R}^{\infty}(\mathcal{N}_{C})$ similar to (\ref{asymdef}). Then there exist random variables $U$ and $V$ satisfying $U\leftrightarrow V\leftrightarrow X\leftrightarrow (Y,Z)$ such that $\mathcal{R}^{\infty}(\mathcal{N}_{C})$ equals the union over all distributions of rate quadruples $(R_{0},R_{1},R_{s},R_{d})$ obeying:
\begin{align*}
R_{0}&\leq\text{min}\big[I(U;Y)_{p},I(U;Z)_{p}\big], \\
R_{0}+R_{1}+R_{s}&\leq I(V;Y|U)_{p}+\text{min}\big[I(U;Y)_{\rho},I(U;Z)_{p}\big], \\
R_{s}&\leq I(V;Y|U)_{p}-I(V;Z|U)_{p}, \\
R_{1}+R_{d}&\geq I(V;Z|U)_{p}+I(X;Z|V)_{p},\\
R_{d}&\geq I(X;Z|V)_{p}.
\end{align*}
\begin{IEEEproof}
This is a simple corollary of Theorem \ref{asyt}. If we assume the channel outputs $B$ and $C$ are classical, then we know that all systems will be simultaneously diagonalizable and the regularization is not needed. Letting $Y\coloneqq B$ and $Z\coloneqq C$ finishes the proof.
\end{IEEEproof}
\end{corollary}
In the following corollary we recover a result for quantum broadcast channel without any secrecy requirement \cite{6034754}.
\begin{corollary}[Theorem in \cite{6034754}]
Consider the quantum broadcast channel $\mathcal{N}^{A\rightarrow BC}$. The capacity region for the transmission of common and private message  $C^{\infty}(\mathcal{N})$ of $\mathcal{N}$ is given as follows\footnote{This is defined similar to (\ref{asymdef}).}:
\begin{align*}
C^{\infty}(\mathcal{N})=\bigcup_{\ell=1}^{\infty}\frac{1}{\ell}C^{\infty}_{1}(\mathcal{N}),
\end{align*}
where $C^{\infty}_{1}(\mathcal{N})$ is the union over all states $\rho^{UVBC}$ arising from the channel, of the rate pairs $(R_{0},R_{1})$ obeying
\begin{align*}
R_{0}&\leq\text{min}\big[I(U;B)_{\rho},I(U;C)_{\rho}\big], \\
R_{0}+R_{1}&\leq I(V;B|U)_{\rho}+\text{min}\big[I(U;B)_{\rho},I(U;C)_{\rho}\big].
\end{align*}
\begin{IEEEproof}
By dropping the secrecy requirement, the rate of the confidential message in Theorem \ref{asyt} will add up to that of the private message. Note that this region is slightly different in appearance compared to the Theorem 1 in \cite{6034754}. However, the discussion leading to the equations (17) and (18) in that paper indicates their equivalence: part (or whole) of the common message may contain information intended for Charlie such that Bob does not have any interest in learning those information; This leads to a slightly different region but the scenario and the rate region are essentially the same in that in superposition coding Bob is supposed to decode the common message in whole and maybe ignore its content afterwards.
\end{IEEEproof}
\end{corollary}
\section{Conclusion}
We have studied the interplay between common, private and confidential messages with rate-limited randomness in the one-shot regime of a quantum broadcast channel. We have proved the optimality of our rate region by finding matching converse bounds. To establish our achievability results, we have proved a conditional version of the convex-split lemma whereby we have shown the channel resolvability problem in the one-shot regime via superpositions. By evaluating our rate regions in the asymptotic i.i.d setting, we recovered several well-known results in the literature.

\appendices
\section{Proof of lemmas}

To prove Lemma \ref{upper1}, we need the following lemma.
\begin{lemma}
\label{ii1}
For quantum states $\rho^{AB}$ and $\sigma^{B}$, there exists a state $\rho'^{A}\in\cB(\rho^{A})$ such that:
\begin{align*}
D_{\ma}(\rho^{AB}\|\rho'^{A}\otimes\sigma^{B})\leq D_{\ma}(\rho^{AB}\|\rho^{A}\otimes\sigma^{B}).
\end{align*}
\begin{IEEEproof}
Trivial.
\end{IEEEproof}
\end{lemma}
\begin{IEEEproof}[Proof of lemma \ref{upper1}]
In the result of Lemma \ref{ii1}, let $\rho^{*AB}$ be the optimizer in the definition of $\widetilde{I}_{\ma}^{\varepsilon}(A;B)_{\rho}$, by substituting this state we will have,
\begin{align*}
D_{\ma}(\rho^{*AB}\|\rho'^{A}\otimes\sigma^{B})\leq D_{\ma}(\rho^{*AB}\|\rho^{*A}\otimes\sigma^{B}).
\end{align*}
Let $\sigma^{B}\coloneqq\rho^{B}$ and choose $\rho'^{A}=\rho^{A}$ (this is possible since $\Pu(\rho^{A},\rho^{*A})\leq\varepsilon$) and then
\begin{align*}
D_{\ma}(\rho^{*AB}\|\rho^{A}\otimes\rho^{B})\leq D_{\ma}(\rho^{*AB}\|\rho^{*A}\otimes\rho^{B}).
\end{align*}
Then the result follows by definitions of the quantities.
\end{IEEEproof}
We need the following lemma to prove Lemma \ref{upper2}.

\begin{lemma}
\label{ii}
For quantum states $\crho$ and $\sigma^{XAB}=\sum_{x}q(x)\ketbra{x}\otimes\sigma^{A}_{x}\otimes\sigma_{x}^{B}$, there exists a state $\rho'^{XAB}\in\mathcal{B}^{\varepsilon}(\rho^{XAB})$ classical on $X$ such that:
\begin{align*}
D_{\ma}\big(\rho'^{XAB}\big\|\sum_{x}p'(x)\ketbra{x}\otimes\rho'^{A}_{x}\otimes\sigma_{B}^{x}\big)\leq D_{\ma}\big(\rho^{XAB}\big\|\sum_{x}q(x)\ketbra{x}\otimes\sigma_{x}^{A}\otimes\sigma_{x}^{B}\big)+\log(\frac{1}{1-\sqrt{1-\varepsilon^{2}}}+1).
\end{align*}
\begin{IEEEproof}
The proof is inspired by \cite{ANR2017} and \cite{CBR14}. Let $\rho^{XABC}$ be a purification of $\rho^{XAB}$ and $\varepsilon>0$. Further let $\Pi^{BC}\in\mathcal{H}_{BC}$ be a projector that is defined as the dual projector of the minimum rank projector $\Pi^{XA}$ with $\text{supp}(\Pi^{XA})\subseteq$ supp$(\rho^{XA})$. The projector $\Pi^{XA}$ is set to minimize $\big\|\Pi^{XA}\Gamma^{XA}\Pi^{XA}\big\|_{\infty}$ while fulfilling $P(\rho^{XABC},\tilde{\rho}^{XABC})\leq\varepsilon$ in which $\Gamma^{XA}\coloneqq(\rho^{XA})^{-\frac{1}{2}}\sigma^{XA}(\rho^{XA})^{-\frac{1}{2}}$ and $\tilde{\rho}^{XABC}\coloneqq\Pi^{BC}\rho^{XABC}\Pi^{BC}$. From Lemma \ref{pur}, we know the following 
\begin{align*}
\Pu(\rho^{XABC},\Pi^{BC}\rho^{XABC}\Pi^{BC})\leq\sqrt{2\Tr\Pi^{BC}_{\perp}\rho-(\Tr\Pi^{BC}_{\perp}\rho)^{2}}=\sqrt{2\Tr\Pi^{XA}_{\perp}\rho-(\Tr\Pi^{XA}_{\perp}\rho)^{2}}.
\end{align*}
If we let $\Tr\Pi^{XA}_{\perp}\rho\leq1-\sqrt{1-\varepsilon^{2}}$, then we will have  $P(\rho^{XABC},\tilde{\rho}^{XABC})\leq\varepsilon$ since $t\mapsto\sqrt{2t-t^{2}}$ is monotonically increasing over $[0,1]$. Now we choose $\Pi^{XA}$ to be the projector onto the smallest eigenvalues of $\Gamma^{XA}$ such that the aforementioned restriction holds, which in turn, results in the minimization of $\big\|\Pi^{XA}\Gamma^{XA}\Pi^{XA}\big\|_{\infty}$. Let $\Pi'^{XA}$ denote the projector onto the largest remaining eigenvalue of $\Pi^{XA}\Gamma^{XA}\Pi^{XA}$. Notice that $\Pi^{XA}$ and $\Pi'^{XA}$ commute with $\Gamma^{XA}$. Then we have the following:
\begin{align*}
\big\|\Pi^{XA}\Gamma^{XA}\Pi^{XA}\big\|_{\infty}=\Tr(\Pi'^{XA}\Gamma^{XA})=\min_{\mu^{XA}}\frac{\Tr(\mu^{XA}\Gamma^{XA})}{\Tr\mu^{XA}},
\end{align*} 
where the minimization is over all operators in the support of $\Pi'^{XA}+\Pi^{XA}_{\perp}$. Choosing $\mu^{XA}=(\Pi'^{XA}+\Pi^{XA}_{\perp})\rho^{XA}(\Pi'^{XA}+\Pi^{XA}_{\perp})$, we will have:
\begin{align*}
\big\|\Pi^{XA}\Gamma^{XA}\Pi^{XA}\big\|_{\infty}\leq\frac{\Tr\{(\Pi'^{XA}+\Pi^{XA}_{\perp})\rho^{XA}(\Pi'^{XA}+\Pi^{XA}_{\perp})\Gamma^{XA}\}}{\Tr\{(\Pi'^{XA}+\Pi^{XA}_{\perp})\rho^{XA}(\Pi'^{XA}+\Pi^{XA}_{\perp})\}}\leq\frac{1}{1-\sqrt{1-\varepsilon^{2}}},
\end{align*}
where from the fact that $\Pi'^{XA}$ and $\Pi^{XA}_{\perp}$ commute with $\Gamma^{XA}$, we have $\Tr\{(\Pi'^{XA}+\Pi^{XA}_{\perp})\rho^{XA}(\Pi'^{XA}+\Pi^{XA}_{\perp})\Gamma^{XA}\}=\Tr\{(\Pi'^{XA}+\Pi^{XA}_{\perp})(\rho^{XA})^{1/2}\Gamma^{XA}(\rho^{XA})^{1/2}\}\leq\Tr\{(\rho^{XA})^{1/2}\Gamma^{XA}(\rho^{XA})^{1/2}\}=\Tr\sigma^{XA}=1$. Moreover, the definition of $\Pi^{XA}$ implies that $\Tr\{(\Pi'^{XA}+\Pi^{XA}_{\perp})\rho^{XA}\}\geq 1-\sqrt{1-\varepsilon^{2}}$.
Let $\gamma\coloneqq D_{max}\big(\rho^{XAB}\big\|\sum_{x}q(x)\ketbra{x}\otimes\sigma_{x}^{A}\otimes\sigma_{x}^{B}\big)$ and $\sigma^{X-B}\coloneqq\sum_{x}\ketbra{x}\otimes\sigma_{x}^{B}$. For state $\tilde{\rho}^{XABC}$ introduced above, we can write:
\begin{align*}
D_{\ma}(\tilde{\rho}^{XAB}&\|\sum_{x}p(x)\ketbra{x}\otimes\rho^{A}_{x}\otimes\sigma_{x}^{B})\\
&=\log\big\|\big(\sum_{x}p(x)\ketbra{x}\otimes\rho^{A}_{x}\otimes\sigma_{x}^{B}\big)^{-\frac{1}{2}}\tilde{\rho}^{XAB}\big(\sum_{x}p(x)\ketbra{x}\otimes\rho^{A}_{x}\otimes\sigma_{x}^{B}\big)^{-\frac{1}{2}}\big\|_{\infty}\\
&=\log\big\|\big(\sum_{x}p(x)\ketbra{x}\otimes\rho^{A}_{x}\otimes\sigma_{x}^{B}\big)^{-\frac{1}{2}}\Tr_{C}\{\Pi^{BC}\rho^{XABC}\Pi^{BC}\}\big(\sum_{x}p(x)\ketbra{x}\otimes\rho^{A}_{x}\otimes\sigma_{x}^{B}\big)^{-\frac{1}{2}}\big\|_{\infty}\\
&=\log\big\|(\sigma^{X-B})^{-\frac{1}{2}}\Tr_{C}\{(\rho^{XA})^{-\frac{1}{2}}\otimes\Pi^{BC}\rho^{XABC}(\rho^{XA})^{-\frac{1}{2}}\otimes\Pi^{BC}\}(\sigma^{X-B})^{-\frac{1}{2}}\big\|_{\infty}\\
&=\log\big\|(\sigma^{X-B})^{-\frac{1}{2}}(\rho^{XA})^{-\frac{1}{2}}\Pi^{XA}\rho^{XAB}(\rho^{XA})^{-\frac{1}{2}}\Pi^{XA}(\sigma^{X-B})^{-\frac{1}{2}}\big\|_{\infty}\\
&\leq\log2^{\gamma}\big\|(\sigma^{X-B})^{-\frac{1}{2}}(\rho^{XA})^{-\frac{1}{2}}\Pi^{XA}(\sum_{x}q(x)\ketbra{x}\otimes\sigma_{x}^{A}\otimes\sigma_{x}^{B})(\rho^{XA})^{-\frac{1}{2}}\Pi^{XA}(\sigma^{X-B})^{-\frac{1}{2}}\big\|_{\infty}\\
&=\log2^{\gamma}\big\|(\rho^{XA})^{-\frac{1}{2}}\Pi^{XA}\sum_{x}q(x)\ketbra{x}\otimes\sigma_{x}^{A}\otimes(\sigma_{x}^{B})^{-\frac{1}{2}}\sigma_{x}^{B}(\sigma_{x}^{B})^{-\frac{1}{2}}(\rho^{XA})^{-\frac{1}{2}}\Pi^{XA}\big\|_{\infty}\\
&=\log2^{\gamma}\big\|(\rho^{XA})^{-\frac{1}{2}}\Pi^{XA}\sum_{x}q(x)\ketbra{x}\otimes\sigma_{x}^{A}\otimes\mathbbm{1}^{B}(\rho^{XA})^{-\frac{1}{2}}\Pi^{XA}\big\|_{\infty}\\
&=\gamma+\log\big\|\Pi^{XA}\Gamma^{XA}\Pi^{XA}\big\|_{\infty}\\
&\leq D_{\ma}\big(\rho^{XAB}\big\|\sum_{x}q(x)\ketbra{x}\otimes\sigma_{x}^{A}\otimes\sigma_{x}^{B}\big)+\log\frac{1}{1-\sqrt{1-\varepsilon^{2}}}.
\end{align*}
Define the positive semi-definite operator $\kappa^{XA}\coloneqq\rho^{XA}-\tilde{\rho}^{XA}$ and Let $\bar{\rho}^{XAB}\coloneqq\tilde{\rho}^{XAB}+\kappa^{XA}\otimes\sigma^{X-B}$.
It can be easily checked that $\bar{\rho}^{XA}=\rho^{XA}$. Moreover, in the following we show that $P(\bar{\rho}^{XAB},\rho^{XAB})\leq\varepsilon$:
\begin{align*}
F(\bar{\rho}^{XAB},\rho^{XAB})&\geq\big\|\sqrt{\tilde{\rho}^{XAB}}\sqrt{\rho^{XAB}}\big\|_{1}+1-\Tr\rho^{XAB}\\
&\geq\big\|\sqrt{\tilde{\rho}^{XABC}}\sqrt{\rho^{XABC}}\big\|_{1}+1-\Tr\rho^{XAB}\\
&=1-\Tr\Pi_{\perp}^{BC}\rho^{BC}\\
&\geq\sqrt{1-\varepsilon^{2}}.
\end{align*} 
The first inequality follows from Lemma \ref{equt} and the fact that by construction $\tilde{\rho}^{XAB}\leq\bar{\rho}^{XAB}$, therefore $\big\|\sqrt{\tilde{\rho}^{XAB}}\sqrt{\rho^{XAB}}\big\|_{1}\leq\big\|\sqrt{\bar{\rho}^{XAB}}\sqrt{\rho^{XAB}}\big\|_{1}$. The second inequality follows from the fact that fidelity is monotonically non-decreasing with respect to CPTP maps. The equality stems from Lemma \ref{equt} and the last inequality is the assumption. And finally from the relation between the purified distance and the fidelity the desired inequality follows. We continue as follows:
\begin{align*}
D_{\ma}\big(\bar{\rho}^{XAB}\big\|\bar{\rho}^{XA}\otimes\sigma^{X-B}\big)&=\log\ct(\bar{\rho}^{XA})^{-\frac{1}{2}}\otimes(\sigma^{X-B})^{-\frac{1}{2}}\bar{\rho}^{XAB}(\bar{\rho}^{XA})^{-\frac{1}{2}}\otimes(\sigma^{X-B})^{-\frac{1}{2}}\ct_{\infty}\\
&=\log\ct(\rho^{XA})^{-\frac{1}{2}}\otimes(\sigma^{X-B})^{-\frac{1}{2}}\bar{\rho}^{XAB}(\rho^{XA})^{-\frac{1}{2}}\otimes(\sigma^{X-B})^{-\frac{1}{2}}\ct_{\infty}\\
&\leq\log\Big(\ct(\rho^{XA})^{-\frac{1}{2}}\otimes(\sigma^{X-B})^{-\frac{1}{2}}\tilde{\rho}^{XAB}(\rho^{XA})^{-\frac{1}{2}}\otimes(\sigma^{X-B})^{-\frac{1}{2}}\ct_{\infty}+1\Big)\\
&\leq\log(2^{\gamma}\frac{1}{1-\sqrt{1-\varepsilon^{2}}}+1)\\
&\leq D_{\ma}\big(\rho^{XAB}\big\|\sum_{x}q(x)\ketbra{x}\otimes\sigma_{x}^{A}\otimes\sigma_{x}^{B}\big)+\log(\frac{1}{1-\sqrt{1-\varepsilon^{2}}}+1),
\end{align*}
where in the first inequality we have used $\bar{\rho}^{XAB}\leq\tilde{\rho}^{XAB}+\rho^{XA}\otimes\sigma^{B}$ and in the final inequality we have used the fact that $2^{\gamma}\geq\Tr\rho^{XAB}=1$. Now similar to Remark \ref{pinch}, a pinching map is applied to the left hand-hand side to conclude from the monotonicity of the max-relative entropy that $X$ system is classical.
\end{IEEEproof}
\end{lemma}
\begin{IEEEproof}[Proof of Lemma \ref{upper2}]
From the result given in Lemma \ref{ii} onward, let $\rho^{*XAB}$ be the optimizer for $D_{max}^{\varepsilon}\big(\rho^{XAB}\big\|\sum_{x}q(x)\ketbra{x}\otimes\sigma_{x}^{A}\otimes\sigma_{x}^{B}\big)$. We argued that this state will be classical on $X$. Then there exists a state $\bar{\rho}^{XAB}\in\cB(\rho^{*XAB})$ classical on $X$ such that
\begin{align*}
D_{max}\big(\bar{\rho}^{XAB}\big\|\sum_{x}\bar{p}(x)\ketbra{x}\otimes\bar{\rho}^{A}_{x}\otimes\sigma_{B}^{x}\big)\leq D_{max}\big(\rho^{*XAB}\big\|\sum_{x}q(x)\ketbra{x}\otimes\sigma_{x}^{A}\otimes\sigma_{x}^{B}\big)+\log(\frac{1}{1-\sqrt{1-\varepsilon^{2}}}+1).
\end{align*}
From the triangle inequality for the purified distance it is seen that $\bar{\rho}^{XAB}\in\mathcal{B}^{2\varepsilon}(\rho^{XAB})$. Choosing $q(x)=p(x), \sigma^{A}_{x}=\rho^{A}_{x},\sigma_{x}^{B}=\rho^{B}_{x}$ for all $x$, finishes the job.
\end{IEEEproof}
To prove Lemma \ref{lower2}, we need to following lemma.
\begin{lemma}
\label{ii2}
Let $\rho^{XAB}$ and $\sigma^{B}$ be a quantum states. There exists a state $\rho'^{XAB}\in\cB(\rho)$ classical on $X$ such that:
\begin{align*}
D_{\ma}(\rho^{XAB}\|\sum_{x}p'(x)\ketbra{x}\otimes\rho'^{A}_{x}\otimes\sigma^{B}_{x})\leq D_{\ma}(\rho^{XAB}\|\sum_{x}p(x)\ketbra{x}\otimes\rho^{A}_{x}\otimes\sigma^{B}_{x}).
\end{align*}  
\begin{IEEEproof}
Trivial.
\end{IEEEproof}
\end{lemma}
\begin{IEEEproof}[proof of Lemma \ref{lower2}]
Let $\rho^{*XAB}$ be the optimizer in the definition of the PSCMMI. By substituting it in Lemma \ref{ii2}, we will have:
\begin{align*}
D_{\ma}(\rho^{*XAB}\|\sum_{x}p'(x)\ketbra{x}\otimes\rho'^{A}_{x}\otimes\sigma^{B}_{x})\leq D_{\ma}(\rho^{*XAB}\|\sum_{x}p^{*}(x)\ketbra{x}\otimes\rho^{*A}_{x}\otimes\sigma^{B}_{x}).
\end{align*} 
Let $\rho'^{XA}=\rho^{XA}$ and $\sigma^{B}=\rho^{B}$. Then the result follows from the definition of the quantities.
\end{IEEEproof}
\begin{IEEEproof}[Proof of Lemma \ref{convexproof}]
Similar to Lemma 11 in \cite{Anurag-convex-split}, the proof follows by straightforward calculation as shown below:
\begin{align*}
&\sum_{i}p(i)\left(D(\rho^{XA}_{i}||\theta^{XA})-D(\rho^{XA}_{i}||\rho^{XA})\right)\\
&\hspace*{2cm}=\sum_{i}p(i)\left(\Tr\{\rho^{XA}_{i}\log\rho^{XA}_{i}\}-\text{tr}\{\rho^{XA}_{i}\log\theta^{XA}\}-\Tr\{\rho^{XA}_{i}\log\rho^{XA}_{i}\}+\Tr\{\rho^{XA}_{i}\log\rho^{XA}\}\right)\\
&\hspace*{2cm}=\Tr\{\sum_{i}p(i)\rho^{XA}_{i}\log\rho^{XA}\}-\Tr\{\sum_{i}p(i)\rho^{XA}_{i}\log\theta^{XA}\}=\Tr\{\rho^{XA}\log\rho^{XA}\}-\Tr\{\rho^{XA}\log\theta^{XA}\}\\
&\hspace*{2cm}=D(\rho^{XA}||\theta^{XA}).
\end{align*}
\end{IEEEproof}
\begin{IEEEproof}[Proof of Lemma \ref{condconex}]
The proof is similar to the proof of its uncontional version \cite{Anurag-convex-split}. For the convenience sake, we let $\sigma^{B_{-j}}_{x}\coloneqq\sigma^{B_{1}}_{x}\otimes...\sigma^{B_{j-1}}_{x}\otimes\sigma^{B_{j+1}}_{x}\otimes...\otimes\sigma^{B_{n}}_{x}$ and $\sigma^{B_{+j}}_{x}\coloneqq\sigma^{B_{1}}_{x}\otimes...\otimes\sigma^{B_{n}}_{x}$. By adopting this notation, we can see that $\tau^{XAB_{1}...B_{n}}=\frac{1}{n}\sum_{j=1}^{n}\sum_{x}p(x)|x\rangle\langle x|^{X}\otimes\rho^{AB}_{x}\otimes\sigma^{B_{-j}}_{x}$. We use Lemma \ref{convexproof} to write the following:
\begin{align}
\nonumber
D\big(\tau^{XAB_{1}...B_{n}}\big\|&\sum_{x}p(x)|x\rangle\langle x|^{X}\otimes\rho^{A}_{x}\otimes\sigma^{B_{+j}}_{x} \big)\\
\label{first}
&\hspace*{3cm}=\frac{1}{n}\sum_{j}D\big(\sum_{x}p(x)|x\rangle\langle x|^{X}\otimes\rho^{AB_{j}}_{x}\otimes\sigma_{x}^{B_{-j}}\big\|\sum_{x}p(x)|x\rangle\langle x|^{X}\otimes\rho^{A}_{x}\otimes\sigma^{B_{+j}}_{x}\big)\\
\label{second}
&\hspace*{3cm}-\frac{1}{n}\sum_{j}D\big(\sum_{x}p(x)|x\rangle\langle x|^{X}\otimes\rho^{AB_{j}}_{x}\otimes\sigma_{x}^{B_{-j}}\big\|\tau^{XAB_{1}...B_{n}}\big).
\end{align} 
From the invariance of the relative entropy with respect to tensor product states, the term inside the summation in (\ref{first}) equals $D\big(\sum_{x}p(x)|x\rangle\langle x|^{X}\otimes\rho^{AB_{j}}_{x}\big\|\sum_{x}p(x)|x\rangle\langle x|^{X}\otimes\rho^{A}_{x}\otimes\sigma^{B_{j}}_{x}\big)$. Besides, from the monotonicity of the quantum relative entropy, by applying $\Tr_{B_{1},...B_{j-1},B_{j+1},...,B_{n}}\{.\}$ to the term inside summation in (\ref{second}), it is lower bounded by $D\big(\sum_{x}p(x)|x\rangle\langle x|^{X}\otimes\rho^{AB_{j}}_{x}\big\|\tau^{XAB_{j}}\big)$ where $\tau^{XAB_{j}}\coloneqq\sum_{x}p(x)|x\rangle\langle x|^{X}\otimes\big(\frac{1}{n}\rho^{AB_{j}}_{x}+(1-\frac{1}{n})(\rho^{A}_{x}\otimes\sigma_{x}^{B_{j}})\big)$. Let $k$ be such that $\rho^{XAB_{j}}\leq2^{k}\sum_{x}p(x)|x\rangle\langle x|^{X}\otimes\rho^{A}_{x}\otimes\sigma_{x}^{B_{j}}$. Therefore, we will have $\rho^{XAB_{j}}\leq(1+\frac{2^{k}-1}{n})\sum_{x}p(x)|x\rangle\langle x|^{X}\otimes\rho^{A}_{x}\otimes\sigma_{x}^{B_{j}}$. Consider the following chain:
\begin{align*}
D\big(\rho^{XAB_{j}}\big\|\tau^{XAB_{j}}\big)=&\Tr\big\{\rho^{XAB_{j}}\log\rho^{XAB_{j}}\big\}-\Tr\big\{\rho^{XAB_{j}}\log\tau^{XAB_{j}}\big\}\\
&\geq \Tr\big\{\rho^{XAB_{j}}\log\rho^{XAB_{j}}\big\}-\Tr\big\{\rho^{XAB_{j}}\log(\sum_{x}p(x)|x\rangle\langle x|^{X}\otimes\rho^{A}_{x}\otimes\sigma_{x}^{B_{j}})\big\}-\log(1+\frac{2^{k}-1}{n})\\
&=D\big(\rho^{XAB_{j}}\big\|\sum_{x}p(x)|x\rangle\langle x|^{X}\otimes\rho^{A}_{x}\otimes\sigma_{x}^{B_{j}}\big)-\log(1+\frac{2^{k}-1}{n}),
\end{align*}
where the inequality comes from the fact that if $A$ and $B$ are positive semidefinite operators and $A\leq B$, then $\log A\leq\log B$.
Plugging the findings above into (\ref{first}) and (\ref{second}) yields:
\begin{align*}
D\big(\tau^{XAB_{1}...B_{n}}\big\|\sum_{x}p(x)|x\rangle\langle x|^{X}\otimes\rho^{A}_{x}\otimes\sigma^{B_{+j}}_{x} \big)&\leq\frac{1}{n}\sum_{j}D\big(\rho^{XAB_{j}}\big\|\sum_{x}p(x)|x\rangle\langle x|^{X}\otimes\rho^{A}_{x}\otimes\sigma_{x}^{B_{j}}\big)\\
&-\frac{1}{n}\sum_{j}D\big(\rho^{XAB_{j}}\big\|\sum_{x}p(x)|x\rangle\langle x|^{X}\otimes\rho^{A}_{x}\otimes\sigma_{x}^{B_{j}}\big)+\log(1+\frac{2^{k}-1}{n})\\
&\leq\log(1+\frac{2^{k}}{n}).
\end{align*}
By choosing $n=\ceil{\frac{2^{k}}{\delta^{2}}}$, it follows that $D\big(\tau^{XAB_{1}...B_{n}}\big\|\sum_{x}p(x)|x\rangle\langle x|^{X}\otimes\rho^{A}_{x}\otimes\sigma^{B_{+j}}_{x} \big)\leq\log(1+\delta^{2})$. From Pinsker's inequality (\ref{Pin}), we also can see that $F^{2}\big(\tau^{XAB_{1}...B_{n}},\sum_{x}p(x)|x\rangle\langle x|^{X}\otimes\rho^{A}_{x}\otimes\sigma^{B_{+j}}_{x} \big)\geq \frac{1}{1+\delta^{2}}\geq 1-\delta^{2}$. From definition of the purified distance, it can be easily seen that $P\big(\tau^{XAB_{1}...B_{n}},\sum_{x}p(x)|x\rangle\langle x|^{X}\otimes\rho^{A}_{x}\otimes\sigma^{B_{+j}}_{x} \big)\leq\delta$.
\end{IEEEproof}
\begin{IEEEproof}[Proof of Corollary \ref{cofcondconvex}]
Let $\tilde{\rho}^{XAB}$ be the optimal state achieving the minimum for $k$. Then from the conditional convex-split lemma we know that:
\begin{align}
\label{one1}
P(\tilde{\tau}^{XAB_{1}...B_{n}},\sum_{x}\tilde{p}(x)|x\rangle\langle x|^{X}\otimes\tilde{\rho}^{A}_{x}\otimes\sigma^{B_{1}}_{x}\otimes...\otimes\sigma^{B_{n}}_{x})\leq\delta,
\end{align} 
where
\begin{align*}
\tilde{\tau}^{XAB_{1}...B_{n}}\coloneqq\sum_{x}\tilde{p}(x)|x\rangle\langle x|^{X}\otimes\big(\frac{1}{n}\sum_{j=1}^{n}\tilde{\rho}^{AB_{j}}_{x}\otimes\sigma^{B_{1}}_{x}\otimes...\otimes\sigma^{B_{j-1}}_{x}\otimes\sigma^{B_{j+1}}_{x}\otimes\sigma^{B_{n}}_{x}\big).
\end{align*}
From the concavity of the fidelity as well as its invariance with respect to tensor product states, the following can be seen:
\begin{align}
\label{two2}
P(\tilde{\tau}^{XAB_{1}...B_{n}},\tau^{XAB_{1}...B_{n}})\leq P(\tilde{\rho}^{XAB},\rho^{XAB})\leq\varepsilon.
\end{align}
Analogously, we have
 \begin{align}
 \label{three3}
 P(\sum_{x}\tilde{p}(x)|x\rangle\langle x|^{X}\otimes\tilde{\rho}^{A}_{x}\otimes\sigma^{B_{1}}_{x}\otimes...\otimes\sigma^{B_{n}}_{x},\sum_{x}p(x)|x\rangle\langle x|^{X}\otimes\rho^{A}_{x}\otimes\sigma^{B_{1}}_{x}\otimes...\otimes\sigma^{B_{n}}_{x})\leq P(\tilde{\rho}^{XA},\rho^{XA})\leq\epsilon.
\end{align} 
Then the desired result is inferred by applying the triangle inequality to (\ref{one1}), (\ref{two2}) and (\ref{three3}).
\end{IEEEproof}

\section*{Acknowledgment}
The first author would like to thank Andreas Winter for being always available to answer his questions and advise him. He is also grateful to Shun Watanabe for walking him through the classical result \cite{Wat-Ooh15} and Marco Tomamichel for useful discussions regarding the entropic quantities appeared in this paper. The work of Farzin Salek and Javier R. Fonollosa is supported by the \enquote{Ministerio de Ciencia, Innovaci\'{o}n y Universidades}, of the Spanish Government, TEC2015-69648-REDC and TEC2016-75067-C4-2-R AEI/FEDER, UE, and the Catalan Government, 2017 SGR 578 AGAUR.

\bibliographystyle{ieeetr}
\bibliography{references.bib}

\begin{thebibliography}{10}

\bibitem{Wyner}
A.~D. Wyner, ``The wire-tap channel,'' {\em The Bell System Technical Journal},
  vol.~54, pp.~1355--1387, Oct 1975.

\bibitem{Csiszar-Korner}
I.~Csiszar and J.~Korner, ``Broadcast channels with confidential messages,''
  {\em IEEE Transactions on Information Theory}, vol.~24, pp.~339--348, May
  1978.

\bibitem{Coverbook}
T.~M. Cover and J.~A. Thomas, {\em Elements of Information Theory (Wiley Series
  in Telecommunications and Signal Processing)}.
\newblock Wiley-Interscience, 2006.

\bibitem{Steinberg-Verdu}
Y.~Steinberg and S.~Verdu, ``Channel simulation and coding with side
  information,'' {\em IEEE Transactions on Information Theory}, vol.~40,
  pp.~634--646, May 1994.

\bibitem{csiszar-korner-book}
I.~Csiszár and J.~Körner, {\em Information Theory: Coding Theorems for
  Discrete Memoryless Systems}.
\newblock Cambridge University Press, 2~ed., 2011.

\bibitem{Bloch-Kliewer}
M.~R. Bloch and J.~Kliewer, ``On secure communication with constrained
  randomization,'' in {\em 2012 IEEE International Symposium on Information
  Theory Proceedings}, pp.~1172--1176, July 2012.

\bibitem{Wat-Ooh15}
S.~Watanabe and Y.~Oohama, ``The optimal use of rate-limited randomness in
  broadcast channels with confidential messages,'' {\em IEEE Transactions on
  Information Theory}, vol.~61, pp.~983--995, Feb 2015.

\bibitem{Chia-ElGamal12}
Y.~K. Chia and A.~E. Gamal, ``Three-receiver broadcast channels with common and
  confidential messages,'' {\em IEEE Transactions on Information Theory},
  vol.~58, pp.~2748--2765, May 2012.

\bibitem{Cai-Winter-Yeung}
N.~Cai, A.~Winter, and R.~W. Yeung, ``Quantum privacy and quantum wiretap
  channels,'' {\em Problems of Information Transmission}, vol.~40,
  pp.~318--336, Oct 2004.

\bibitem{Dev-private}
I.~Devetak, ``The private classical capacity and quantum capacity of a quantum
  channel,'' {\em IEEE Transactions on Information Theory}, vol.~51,
  pp.~44--55, Jan 2005.

\bibitem{Devetak-Shor}
I.~Devetak and P.~W. Shor, ``The capacity of a quantum channel for simultaneous
  transmission of classical and quantum information,'' {\em Communications in
  Mathematical Physics}, vol.~256, pp.~287--303, Jun 2005.

\bibitem{Marcobook}
M.~Tomamichel, {\em Quantum information processing with finite resources :
  mathematical foundations}.
\newblock SpringerBriefs in mathematical physics ; v. 5, 2016.

\bibitem{Renner-single-serving}
R.~Renner, S.~Wolf, and J.~Wullschleger, ``The single-serving channel
  capacity,'' in {\em 2006 IEEE International Symposium on Information Theory},
  pp.~1424--1427, July 2006.

\bibitem{Milan-Nilanjana}
M.~Mosonyi and N.~Datta, ``Generalized relative entropies and the capacity of
  classical-quantum channels,'' {\em Journal of Mathematical Physics}, vol.~50,
  no.~7, p.~072104, 2009.

\bibitem{Wang-Renner}
L.~Wang and R.~Renner, ``One-shot classical-quantum capacity and hypothesis
  testing,'' {\em Phys. Rev. Lett.}, vol.~108, p.~200501, May 2012.

\bibitem{Mark2017}
M.~M. Wilde, ``Position-based coding and convex splitting for private
  communication over quantum channels,'' {\em Quantum Information Processing},
  vol.~16, p.~264, Sep 2017.

\bibitem{ANR2017}
A.~{Anshu}, R.~{Jain}, and N.~A. {Warsi}, ``{One shot entanglement assisted
  classical and quantum communication over noisy quantum channels: A hypothesis
  testing and convex split approach},'' {\em ArXiv e-prints},
  p.~arXiv:1702.01940, Feb. 2017.

\bibitem{Anurag-convex-split}
A.~Anshu, V.~K. Devabathini, and R.~Jain, ``Quantum communication using
  coherent rejection sampling,'' {\em Phys. Rev. Lett.}, vol.~119, p.~120506,
  Sep 2017.

\bibitem{JPN2017}
J.~{Radhakrishnan}, P.~{Sen}, and N.~A. {Warsi}, ``{One-Shot Private Classical
  Capacity of Quantum Wiretap Channel: Based on one-shot quantum covering
  lemma},'' {\em ArXiv e-prints}, p.~arXiv:1703.01932, Mar. 2017.

\bibitem{Covering.A.W.2002}
R.~Ahlswede and A.~Winter, ``Strong converse for identification via quantum
  channels,'' {\em IEEE Transactions on Information Theory}, vol.~48,
  pp.~569--579, March 2002.

\bibitem{Renes-Renner11}
J.~M. Renes and R.~Renner, ``Noisy channel coding via privacy amplification and
  information reconciliation,'' {\em IEEE Transactions on Information Theory},
  vol.~57, pp.~7377--7385, Nov 2011.

\bibitem{Buscemi-Datta-2010}
F.~Buscemi and N.~Datta, ``The quantum capacity of channels with arbitrarily
  correlated noise,'' {\em IEEE Transactions on Information Theory}, vol.~56,
  pp.~1447--1460, March 2010.

\bibitem{Farzin2018}
F.~Salek, A.~Anshu, M.~Hsieh, R.~Jain, and J.~R. Fonollosa, ``One-shot capacity
  bounds on the simultaneous transmission of public and private information
  over quantum channels,'' in {\em 2018 IEEE International Symposium on
  Information Theory (ISIT)}, pp.~296--300, June 2018.

\bibitem{Farzin.arXiv.2018}
F.~{Salek}, A.~{Anshu}, M.-H. {Hsieh}, R.~{Jain}, and J.~R. {Fonollosa},
  ``{One-shot Capacity bounds on the Simultaneous Transmission of Classical and
  Quantum Information},'' {\em ArXiv e-prints}, p.~arXiv:1809.07104, Sept.
  2018.

\bibitem{Anurag-Hayashi-Warsi.2018}
A.~{Anshu}, M.~{Hayashi}, and N.~A. {Warsi}, ``{Secure communication over fully
  quantum Gel'fand-Pinsker wiretap channel},'' {\em arXiv e-prints},
  p.~arXiv:1801.00940, Jan. 2018.

\bibitem{6034754}
J.~{Yard}, P.~{Hayden}, and I.~{Devetak}, ``Quantum broadcast channels,'' {\em
  IEEE Transactions on Information Theory}, vol.~57, pp.~7147--7162, Oct 2011.

\bibitem{Markbook}
M.~M. Wilde, {\em Quantum Information Theory}.
\newblock New York, NY, USA: Cambridge University Press, 1st~ed., 2013.

\bibitem{Csiszar67}
I.~Csisz{\'a}r, ``{Information-type measures of difference of probability
  distributions and indirect observations},'' {\em Studia Sci. Math. Hungar.},
  vol.~2, pp.~299--318, 1967.

\bibitem{20.500.11850/153605}
M.~Tomamichel, {\em A framework for non-asymptotic quantum information theory}.
\newblock PhD thesis, ETH Zurich, 2012.
\newblock Diss., EidgenÃ¶ssische Technische Hochschule ETH ZÃŒrich, Nr.
  20213.

\bibitem{5961850}
M.~{Tomamichel}, C.~{Schaffner}, A.~{Smith}, and R.~{Renner}, ``Leftover
  hashing against quantum side information,'' {\em IEEE Transactions on
  Information Theory}, vol.~57, pp.~5524--5535, Aug 2011.

\bibitem{2018arXiv180607276S}
P.~{Sen}, ``{Inner bounds via simultaneous decoding in quantum network
  information theory},'' {\em arXiv e-prints}, p.~arXiv:1806.07276, Jun 2018.

\bibitem{Nilanjana-2009}
N.~Datta, ``Min- and max-relative entropies and a new entanglement monotone,''
  {\em IEEE Transactions on Information Theory}, vol.~55, pp.~2816--2826, June
  2009.

\bibitem{2012arXiv1211.3141D}
F.~{Dupuis}, L.~{Kraemer}, P.~{Faist}, J.~M. {Renes}, and R.~{Renner},
  ``{Generalized Entropies},'' {\em arXiv e-prints}, p.~arXiv:1211.3141, Nov
  2012.

\bibitem{6574274}
M.~{Tomamichel} and M.~{Hayashi}, ``A hierarchy of information quantities for
  finite block length analysis of quantum tasks,'' {\em IEEE Transactions on
  Information Theory}, vol.~59, pp.~7693--7710, Nov 2013.

\bibitem{li2014}
K.~Li, ``Second-order asymptotics for quantum hypothesis testing,'' {\em Ann.
  Statist.}, vol.~42, pp.~171--189, 02 2014.

\bibitem{Hayashi-Nagaoka-2003}
M.~Hayashi and H.~Nagaoka, ``General formulas for capacity of classical-quantum
  channels,'' {\em IEEE Transactions on Information Theory}, vol.~49,
  pp.~1753--1768, July 2003.

\bibitem{Gamal:2012:NIT:2181143}
A.~E. Gamal and Y.-H. Kim, {\em Network Information Theory}.
\newblock New York, NY, USA: Cambridge University Press, 2012.

\bibitem{Mark-app}
H.~Qi, Q.~Wang, and M.~M. Wilde, ``Applications of position-based coding to
  classical communication over quantum channels,'' {\em Journal of Physics A:
  Mathematical and Theoretical}, vol.~51, no.~44, p.~444002, 2018.

\bibitem{Gentlem-1999}
A.~Winter, ``Coding theorem and strong converse for quantum channels,'' {\em
  IEEE Transactions on Information Theory}, vol.~45, pp.~2481--2485, Nov 1999.

\bibitem{CBR14}
N.~Ciganović, N.~J. Beaudry, and R.~Renner, ``Smooth max-information as
  one-shot generalization for mutual information,'' {\em IEEE Transactions on
  Information Theory}, vol.~60, pp.~1573--1581, March 2014.

\end{thebibliography}

\end{document}